%% file: paper.tex
\documentclass[12pt]{iopart}

\usepackage{iopams}

\usepackage{lscape}

\usepackage[mathscr]{eucal}
\usepackage{array}
\usepackage{epsfig}
\usepackage{psfrag} 
\usepackage{subfigure}
\usepackage{graphicx}
\usepackage{bm}
\usepackage{pifont}

\usepackage{cleveref}

\usepackage{algpseudocode}
\usepackage{algpseudocode}

\newcommand{\eqref}[1]{(\ref{#1})}
\newcommand{\figref}[1]{Fig.~\ref{#1}}
\newcommand{\Figref}[1]{Fig.~\ref{#1}}
\newcommand{\secref}[1]{Section~\ref{#1}}

\usepackage{afterpage}

\input{./symbols_and_definitions.tex}

\usepackage[usenames, dvipsnames]{color}
\usepackage[normalem]{ulem} 

\newcommand{\Replace}[2]{\bgroup\noindent\textcolor{red}{\xout{#1} #2}\egroup\ignorespacesafterend}
\newcommand{\Delete} [1]{\bgroup\noindent\textcolor{red}{\xout{#1}}\egroup\ignorespacesafterend}
\newcommand{\Insert} [1]{\bgroup\noindent\textcolor{blue}{#1}\egroup\ignorespacesafterend}
\newcommand{\MyComment}[1]{\definecolor{Mygray}{gray}{0.50}\bgroup\color{Mygray}\noindent#1\egroup\ignorespacesafterend\quad\hbox{}}

\newcommand \Stefan [1] {\bgroup\noindent[\textcolor{red}{\textbf{Stefan}: #1}]\egroup\ignorespacesafterend}
\newcommand \David[1] {\bgroup\noindent[\textcolor{blue}{\textbf{David}: #1}]\egroup\ignorespacesafterend}
\newcommand \Zoe[1] {\bgroup\noindent[\textcolor{JungleGreen}{\textbf{Zoe}: #1}]\egroup\ignorespacesafterend}

\newcommand{\AlgoComment}[1]{\definecolor{Mygray}{gray}{0.50}\bgroup\color{Mygray}\noindent\hbox{}\hfill\##1\egroup\ignorespacesafterend\quad\hbox{}}

\begin{document}

\title[Avalanches, loading and finite size effects in 2D amorphous plasticity]%
      {Avalanches, loading and finite size effects in 2D amorphous plasticity: results from a finite element model }

\author{Stefan Sandfeld$^1$, Zoe Budrikis$^2$, Stefano Zapperi$^{2, 3,4}$, David Fernandez Castellanos$^1$}
\address{$^1$Institute of Materials Simulation, University of Erlangen-N\"urnberg, Dr.-Mack-Stra{\ss}e 77, 90762 F\"urth, Germany}
\address{$^2$ISI Foundation, Via Alassio 11/c, 10126 Torino, Italy}
\address{$^3$CNR-IENI, Via R. Cozzi 53, 20125 Milano, Italy}
\address{$^4$ Department of Applied Physics, 
Aalto University, P.O. Box 14100, FIN-00076 Aalto, Espoo, Finland}
\ead{stefan.sandfeld@fau.de}

\begin{abstract}
Crystalline plasticity is strongly interlinked with dislocation mechanics and nowadays is relatively well understood. Concepts and  physical models of plastic deformation in amorphous materials on the other hand --- where the concept of linear lattice defects is not applicable --- still are lagging behind.
We introduce an eigenstrain-based finite element lattice model for  simulations of shear band formation and strain avalanches. Our model allows us to study the influence of surfaces and finite size effects on the statistics of avalanches. We find that even with relatively complex loading conditions and open boundary conditions, critical exponents describing avalanche statistics are unchanged, which validates the use of simpler scalar lattice-based models to study these phenomena.
\end{abstract}


\section{Introduction}
\label{sec:intro}
Under mechanical loading, amorphous materials  such as bulk metallic glasses (BMG) and binary particle mixtures exhibit a rich variety of collective phenomena such as strain localization into shear bands and power-law distributed strain avalanches.
Although at the particle level amorphous materials have a disordered structure like a fluid, they nonetheless show a distinct solid-like yielding behaviour. 
A key challenge for understanding the deformation behaviour of such systems is the link between the microscopic and the macroscale behaviour: on the atomic length scale, deformations of BMGs have been widely studied by means of molecular dynamic simulations~\cite{Kobayashi1980, Srolovitz1983, Malandro1998, Malandro1999, Lacks2001}. 
However, molecular dynamic simulations are severely limited in terms of size and time scale, in particular  if predictions are to be compared with experimental results on larger length and time scales together with realistic strain rates. 

A different class of models are mesoscopic models which operate with resolutions well above atomic distances but still aim to capture relevant features of the microstructure and its evolution. Among such models~\cite{Baret2002,Vandembroucq2011,Talamali2011,Talamali2012, Budrikis2013}, a common approach is to represent an amorphous solid undergoing plastic deformation through two competing mechanisms: (i) mutual interaction of localized regions of plastic rearrangements (shear transformations; hereafter STs) through long-range stress fields, and (ii) disorder in form of a fluctuating distribution of local yield stresses. 
An advantage of this approach is that  the same two competing mechanisms can be found also in interface depinning problems for which a number of solution and analysis strategies are available. 
In this spirit, plastic yielding can be understood as a depinning phase transition for which the universality class can be determined through the exponents of power-laws that characterize the distribution of avalanche sizes. 

Determining the universality class has proved to be non-trivial:  although the interactions are long-range and thus one would na\"ively expect mean field behaviour (e.g., in analogy to dislocation systems \cite{Zaiser2006}),  behaviour inconsistent with the mean field universality class has been observed as a consequence of the anisotropy of interactions~\cite{Talamali2011,Budrikis2013,Salerno2013}. To date, studies of depinning models have focused mainly on minimalistic models which treat  localized plastic rearrangements essentially as a point-like phenomenon; periodic boundary conditions (PBCs) are employed for obtaining the resulting stress fields since other boundary conditions drastically increase the complexity of evaluating elastic interaction kernel functions.  While such models are well able to capture many qualitative features of experimentally obtained results, it is in particular the use of PBCs that prevents a detailed comparison with small-scale samples where surface effects become more pronounced. Indeed, it is not {\it a priori} clear that surface effects, which include spatial nonuniformity in external loading, should not affect the universality of the model. 

We therefore take an alternative approach: our mesoscopic model for athermal amorphous plasticity mimics the dynamics of lattice-based models~\cite{Talamali2011,Budrikis2013} but is based on the finite element method (FEM) for calculations of externally applied as well as resulting internal stresses. This allows us to examine how general boundary and surface/loading conditions affect plastic deformation in this class of depinning models, in addition to studying avalanche statistics. 
FEM models have previously been used to study the evolution of STs at finite temperature~\cite{Homer2009, Homer2010, wang2013,Homer2014}; with our FEM model we try to close the gap between FEM models rather used  from the engineering community and modelling approaches which have their roots in the statistical mechanics community, and to this end we present a detailed benchmark of our FEM model with the aforementioned lattice model \cite{Budrikis2013}. The key results of this work are from our study of size- and surface effects in non-periodic situations and their impact on the scaling behaviour. We find that while surfaces have a dramatic effect on observed plastic strain patterns, they do not affect the universal behaviour of amorphous plasticity and we measure critical exponents in agreement with those previously observed in lattice-based models.

This paper is organized as follows: in \secref{sec:models}, we recapitulate the main features of the reference model and introduce the FEM approach taken in our model. As a fundamental comparison of the two models in a static situation we study  in \secref{sec:eshelby} the resulting stresses for an Eshelby inclusion-type of problem. Subsequently, in \secref{sec:PBCvsFree} 
we study how both models perform in complex, time-dependent situations in terms of avalanches and scaling behaviour and investigate finite size effects for both models in periodic configurations, as well as size effects and surface effects for the FEM model in non-periodic situations.

\section{The models}
\label{sec:models}

The essential approach taken in mesoscale descriptions of amorphous plasticity  is to consider plastic activity in  localized units, known as shear transformations (STs), see e.g., \cite{Argon1979101, Argon1983499}. Microscopically, a ST corresponds to a rearrangement of atoms within the bulk as a consequence of local shear stress. It  represents, however, no unique or well-defined volume on an atomic scale. This atomic rearrangement in turn induces a stress state in which the local shear stress again is driving the local deformation. 
In the following we introduce the two models used for our comparisons: the lattice-based model and the finite element model.

\subsection{The reference model}
\label{sec:refmodel}
\begin{figure}[htb]
\hbox{}
\subfigure[]
{\quad\includegraphics[viewport=0 80 484 563, clip, width=0.25\textwidth]{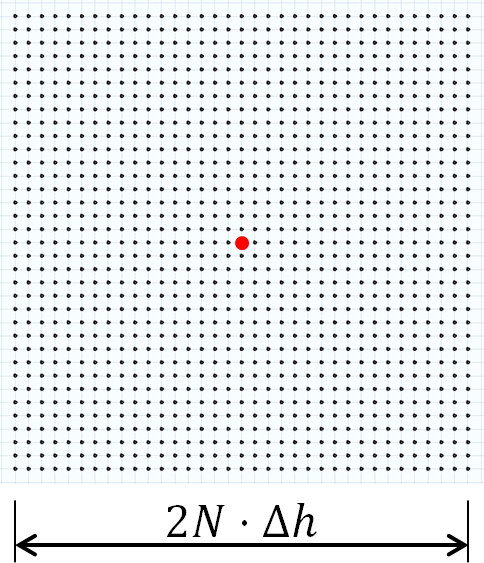}\quad}%
\hfill
\subfigure[]
{\quad\includegraphics[viewport=0 80 465 620, clip, width=0.25\textwidth]{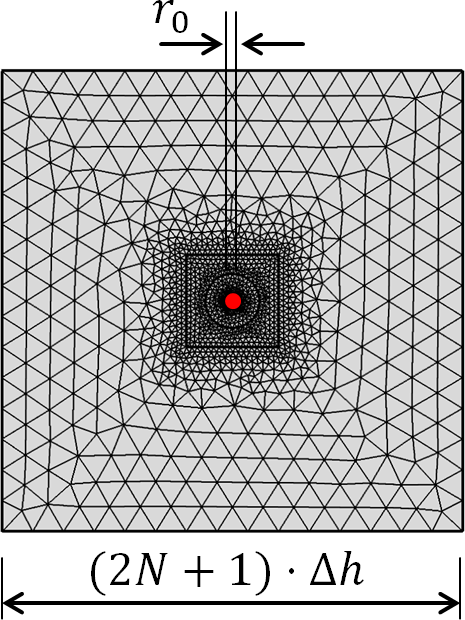}\quad}%
\hfill
\subfigure[]%
{\quad\includegraphics[viewport=0 80 470 631, clip, width=0.25\textwidth]{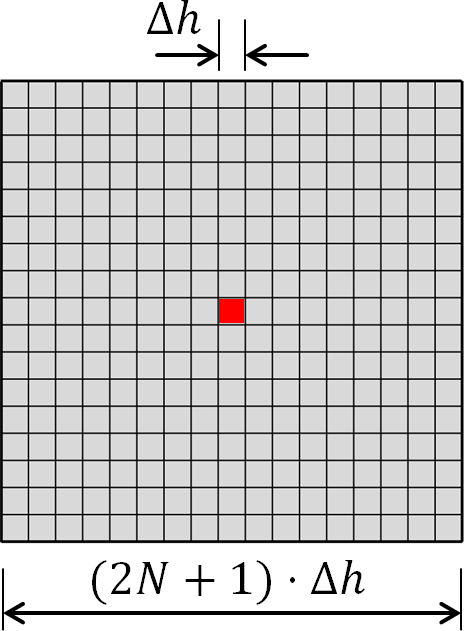}\quad}%
\hbox{}

\caption{\label{fig:system_sketch}
Sketch of the three investigated systems, each of which represents an Eshelby inclusion: (a) shows the regular lattice of points for the reference model; (b) shows the optimized  FEM mesh, which is refined around the small, circular inclusion and which for $r_0\rightarrow 0$ approaches the analytical Eshelby inclusion solution; (c) shows the coarser discretization  used in the simulations in \secref{sec:PBCvsFree}; each  ST is associated with one quadratic element of size $\Delta h$, and the stress is averaged over each element.}
\end{figure}

A common modelling approach is to idealize STs as point-like objects and divide the material into a regular lattice-like structure. The properties of each of the cells are then represented by the point-like STs (as sketched in \figref{fig:system_sketch}(a)).  This fixed structure makes it possible to examine how strain localizations result from elastic interactions between STs \cite{Baret2002, maloney2006, Bailey2007,Talamali2011, Vandembroucq2011, Budrikis2013}. Such lattice-based models typically use periodic boundary conditions, as in general the Green's function describing the elastic interaction stresses is tractable only in infinite or periodic media. Additionally, almost all these models are scalar models in the sense that they do not consider the full stress/strain tensor but rather only the (scalar) shear component.

As a reference model we use the two-dimensional lattice model used by Budrikis and Zapperi \cite{Budrikis2013} to simulate quasi-static loading of a specimen in a shear deformation situation. This model is essentially the same as that developed by Talamali {\it et al}~\cite{Talamali2011, Talamali2012}, but operates with an adiabatically-increasing external drive rather than using extremal dynamics.
An important consequence of adiabatic driving is that the reference model is well-suited to study sizes and durations of strain avalanches, which can be uniquely identified from the evolution of plastic activity. 

In the reference model, the stress acting on each site is the sum of the externally applied uniform shear stress and the internal shear stresses resulting from the plastic deformation of every other site in the system. The evolution proceeds as follows:
 (i) the system is initialized with the yield stress at each point drawn from a uniform random distribution;
 (ii) the external shear stress is increased so that a single site yields as consequence of the local shear stress; 
 (iii) the system relaxes by simultaneously shear-transforming the sites for which the local shear stress is higher than the randomly prescribed yield stress (new yield stresses are given to these positions); 
 (iv) the changes in stresses due to plastic shear strain increments are taken to occur instantaneously, and the local shear stress is recalculated at every point; 
 (v) steps (iii)-(iv) are repeated until the system reaches the equilibrium (i.e., nothing transforms) and then the algorithm returns to step (ii). In the simulations presented in Ref.~\cite{Budrikis2013}, local plastic strains were only allowed to increase (that is, negative stresses had no effect). In the simulations presented here, negative plastic shear strain increments are allowed if the local shear stress is sufficiently negative. In practice, avalanches of net negative shear strain only occur at early times in the simulation as the external drive biases the stresses to be positive, and little difference is seen between the two yielding conditions.

In this model, the stress redistribution mentioned in step (iii) is carried out by a pre-calculated Green's function. In an infinite system, the Green's function is given by $K(\Br)\propto \cos(4\theta)/r^2$. In order to use this kernel in a finite-size simulation, periodic boundary conditions  are imposed. Two possible methods for achieving this are summing over images, or discretizing the  Green's function in Fourier space and obtaining the real space kernel by discrete Fourier transform~\cite{Talamali2011}. Regardless of the used  periodization method, the resulting interaction kernel deviates from its infinite-system behaviour close to the system boundaries. Furthermore, although the `image sum' periodizing approach retains the short-range behaviour of the infinite system quite well, it was previously found that small variations in short-range interactions on the lattice can have strong effects on strain localization and the size distribution of avalanches for small external stresses. However, it should be emphasized that for the purposes of determining the properties of the depinning phase transition, the kernels are equivalent since the universality class of the transition does not depend on short-range interactions.

\subsection{The finite element model}
\label{sec:eistrain}
Finite element simulations of STs take a different approach.  STs within a deforming material are approximated as two-dimensional elements of finite size. A plastic event is introduced into the system by adding a shear \emph{eigenstrain}~\cite{Eshelby1957} increment to the element undergoing a ST, where eigenstrain denotes the stress-free strain that arises as a consequence of inelastic deformation such as a plastic displacement, thermal expansion etc. We refer to \emph{inclusion} as a subset -- in our model, coinciding with a single element -- of the system, that is undergoing such a deformation. 
The actual strain state of an inclusion within the system is different from the eigenstrain due to the constricting effects of the surrounding. This mismatch introduces a stress field, which is the direct consequence of this localized inelastic deformation and which may again cause plastic deformation. Such simulations are able to replicate the mechanical behaviour of BMGs at relevant temperature and stress conditions (hundreds of Kelvin and MPa) \cite{Homer2009, Homer2010, Homer2014}. FEM models have also allowed for studying mechanical properties based on surface effects and external stress gradients such as indentation~\cite{Ai2008} or surface roughness under deformation~\cite{Sandfeld2014} and are applicable in situations with non-trivial boundary conditions. Our eigenstrain-based FEM model implements the same algorithm as the scalar (reference) model and is set up in the spirit of continuum modelling of internal stresses of dislocation systems~\cite{Zaiser2005,Sandfeld2013}. As a preparation we now briefly introduce relevant equations and notations for the general  FEM solution of a solid mechanical problem with eigenstrains.

\subsection{The finite element method for solving linear elasticity boundary value problems. }
\label{sec:FEM}
Linear elastic behaviour of a specimen $\calV\subset\IR^2$ undergoing mechanical deformations can be described by the following set of equations: given a displacement field $\Bu(\Br)=u_i(\Br)$, 
where $\Br\in\calV$ is a point belonging to the specimen and $u_{i=1,2}$ are the components of the displacement field $\Bu(\Br)$, we obtain the infinitesimal strain tensor $\Bve(\Br)=\ve_{ij}(\Br)$ as 
\begin{equation} \label{eq:kinematic}
\ve_{ij} = \frac{1}{2}\left( \frac{\partial u_i}{\partial x_j} + \frac{\partial u_j}{\partial x_i}\right)
\end{equation}
For brevity, we drop the point of evaluation $\Br$, and assume that the Einstein summation convention is applied for double indices. Equilibrium of the solid body \calV in the presence of eigenstrains $\Bve^\ast$ and without body forces is governed by the balance of momentum equation
\begin{equation} \label{eq:BE}
\frac{\partial \sigma_{ij}}{\partial x_j} + \frac{\partial}{\partial x_j}\IC_{ijkl}\ve^\ast_{kl}= 0,
\end{equation}
where $\sigma_{ij}$ are the components of the Cauchy stress tensor and $\IC_{ijkl}$ are the components of the elasticity tensor. Material behaviour is described by the {constitutive equation} which relates stresses and strains
%
%

\begin{equation} \label{eq:CE}
\sigma_{ij}=\IC_{ijkl} \ve_{kl},
\end{equation}

where in our model we assume the tensor of isotropic homogeneous media.
 Using these equations together with the symmetry of $\IC$  one can derive an 
 expression in which only the $u_{i}$ are unknown:
 \begin{equation}\label{eq:D}
 \IC_{ijkl} u_{k,li}+\frac{\partial}{\partial x_j}\IC_{ijkl}\ve^\ast_{kl}=0
 \end{equation}
In general these equations need to be complemented by boundary conditions (BCs), which are prescribed on the surface $\partial\calV$ of the body. Those can be either displacement (`Dirichlet') BCs or traction (`Neumann') BCs.  FEM numerically approximates the solution of \eqref{eq:D} under consideration of those BCs  by discretizing the whole domain into finite-sized non-overlapping elements  defined by a set of interconnected nodes. The solution is based on the so-called weak form which mathematically relaxes the point-wise exact validity of \eqref{eq:D} and the BCs. Finally, the discretized weak form can be solved as a linear system of equations yielding as solution the displacements of the nodes that define the elements. Stresses and strains can be obtained at arbitrary points $\Br$ in a postprocessing step from \eqref{eq:kinematic} and \eqref{eq:CE} together with a suitable interpolation scheme, the 'shape functions'.


Additional care is required with periodic FEM systems: because periodic FEM systems identify the \emph{displacements} of  nodes of opposite surfaces with each other and \emph{not} strains, the  strains  obtained from the solution of the eigenstrain problem need to be corrected~\cite{NematNasser}. The strain $\ve^{\rm FEM}$ obtained from the FEM solver must be modified with the average of the prescribed average eigenstrain $\langle \ve^\ast\rangle$,
\begin{equation}\label{eq:PBCcorr}
	\ve_{12}(\Br) = \ve_{12}^{\rm FEM}(\Br) +  \langle \ve_{12}^\ast\rangle,
	 \qquad \textrm{where}\;\;
 	\langle \ve_{12}^\ast\rangle = \frac{1}{|\calV |}\lint{}{\calV} \ve_{12}^*(\Br) \,\totdiff\calV.
\end{equation}
For a shear eigenstrain the normal components $\varepsilon_{ii}$ are not affected.

\section{A static benchmark of the two models:  the Eshelby inclusion}
\label{sec:eshelby}

One cornerstone of the two models is the correct representation of internal shear stresses $\tauint$ and internal shear strains $\epsint$ which are caused by STs (where we denote by $\tauint$ and $\epsint$ the resulting shear stress/strain from solution of an eigenstrain problem). Thus, our first  test is concerned with the correct representation of a single Eshelby inclusion that we position in the center of our computational domain.
In a dynamic model with  stress redistribution as investigated later, one of course has to calculate the resulting stresses and strains for a large number of STs simultaneously. For benchmarking purposes, though, we compare the result for only one inclusion. Since the problem is linear the result can simply be transferred to more complex systems by superposition.

\begin{figure}
	\centering
	\subfigure[Plots of $\tauint$ along the horizontal $y=0$ and diagonal $x=y$ direction]{%
	\includegraphics[width=0.47\textwidth]{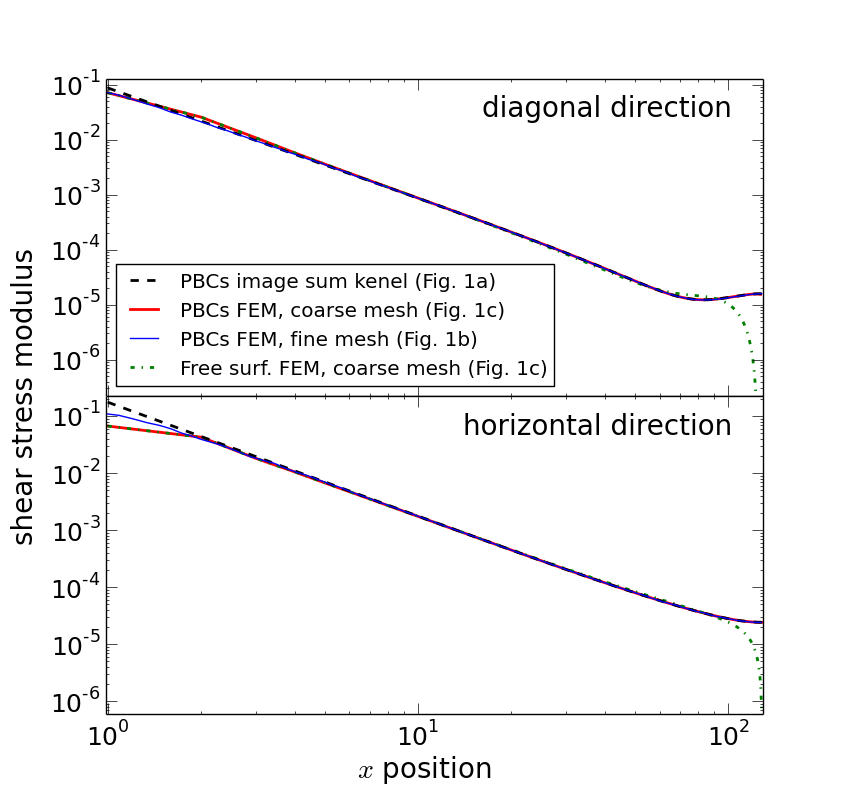}}
	\hfill%
	\subfigure[Absolute value of the difference between $\tauint$ for the systems of %
	\figref{fig:system_sketch}(a) and \ref{fig:system_sketch}(c) %
	 (showing a symmetric quarter of the system)]{%
	\includegraphics[width=0.47\textwidth]	{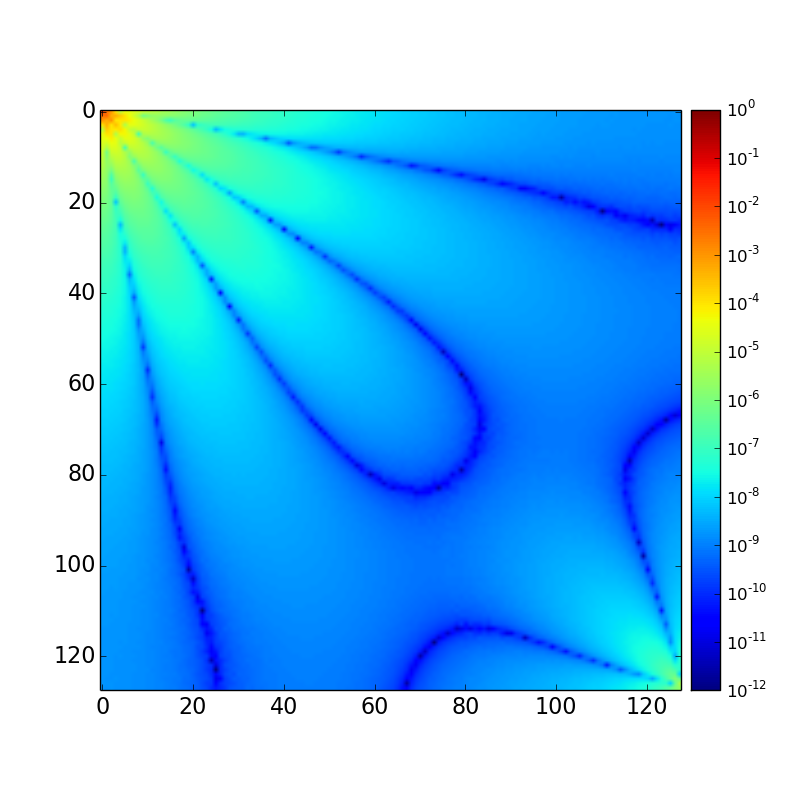}	}
	\caption{\label{fig:eshelby1d} 
	Comparison of shear stress of an Eshelby inclusion 
	for different numerically approximated systems (see \figref{fig:system_sketch}). The main differences arise in the nearest neighbours of the inclusion and at the corners with a very small parasitic stress as a consequence of the FEM periodic boundary implementation. }
\end{figure}

In the reference model a ST is a point-like inclusion  (\figref{fig:system_sketch}(a)) for which  a Green's function yields the resulting internal stress for all lattice points. The FEM model, on the other hand, relies on the approximate computation of the strain field around an inclusion of finite size. We study the effect of two different FEM discretization approaches on the elastic interactions and additionally a non-periodic FEM system:
\begin{itemize}
\item[(i)] The FEM mesh is locally refined (\figref{fig:system_sketch}(b)) consisting of triangular elements using quadratic shape functions. 
In the limit case of a vanishing inclusion size $r_0$, the resulting stress and strain field  should converge to the Green's function for a point inclusion.
\item[(ii)] The FEM mesh is significantly coarser and very regular (\figref{fig:system_sketch}(c)) with quadratic elements, size $\Delta h$ and linear shape functions. The value of stress is averaged over each element and therefore it has the same geometry as the lattice structure of the reference model.  
\item[(iii)] The FEM mesh is the same as in (ii) but now we leave the surfaces free, i.e., the system can deform non-periodically and surfaces alter the internal stress state.
\end{itemize}

\Figref{fig:eshelby1d}(a) shows $\tauint$ plotted along $y=0$ and $x=y$. Obviously, a smaller inclusion results in a more accurate strain field in the vicinity of the inclusion. In the long range, all FEM stresses perfectly  match the reference solution, whereas without the PBC correction \eqref{eq:PBCcorr} large deviations would occur. In the short range close to the center of the inclusion, however, the stress obtained from FEM is significantly smaller. This effect is more pronounced for the coarser mesh and is more obvious for the plot along the horizontal direction $y=0$. Additionally, for the coarse square mesh we averaged the stress for each cell while for the triangular mesh we used the quadratic FEM shape functions for interpolation giving a higher accuracy but at the price of a high computational cost.

The absolute differences  between the Green's function for the reference model and the coarse mesh FEM kernel are shown in \figref{fig:eshelby1d}(b).  The main differences are found close to the inclusion and in the corners of the system. The latter is an artifact arising as a consequence of parasitic stresses introduced by the specific implementation of PBCs in FEM (we have to pin one node to obtain a unique solution). However, this difference is of the order of $10^{-6}$ and is not expected to introduce any appreciable bias in the emerging statistical properties of the system. The difference in the short range, i.e., the nearest neighbours, are larger and  can affect the pattern of strain localization~\cite{Budrikis2013}.

Finally, using again the mesh shown in \figref{fig:system_sketch}(c), we take a look at a system with free surfaces with an inclusion in the center, also plotted in \figref{fig:eshelby1d}(a). The short-range interactions are unchanged (with respect to the same mesh under PBCs) but the long-range is dramatically changed by the free surfaces as compared to the periodic systems: the stress drops to zero because the balance equation ($\div\Bsigma=0$) must be fulfilled for the (traction-free) surface as well as for the bulk.  In the next section we will study dynamically evolving systems and will see if this affects the nature of the depinning transition.

\section{Avalanches, strain localizations and finite size effects}
\label{sec:PBCvsFree} 
We now study the evolution of the FEM model with the coarse, quadratic mesh, \figref{fig:system_sketch}(c), and compare it with the reference model. We start with a FEM model that mimics the reference model as closely as possible (PBCs with a constant external stress) and proceed towards models that better represent realistic physical systems (surfaces together with external stress obtained from FEM, see table \ref{tab:models}).

\begin{figure}[ht]
\centering
\hbox{}\hfill\qquad
 \subfigure[Pure shear ($\tauext=\rm const$)]%
 {\qquad\includegraphics[height=0.22\textwidth]{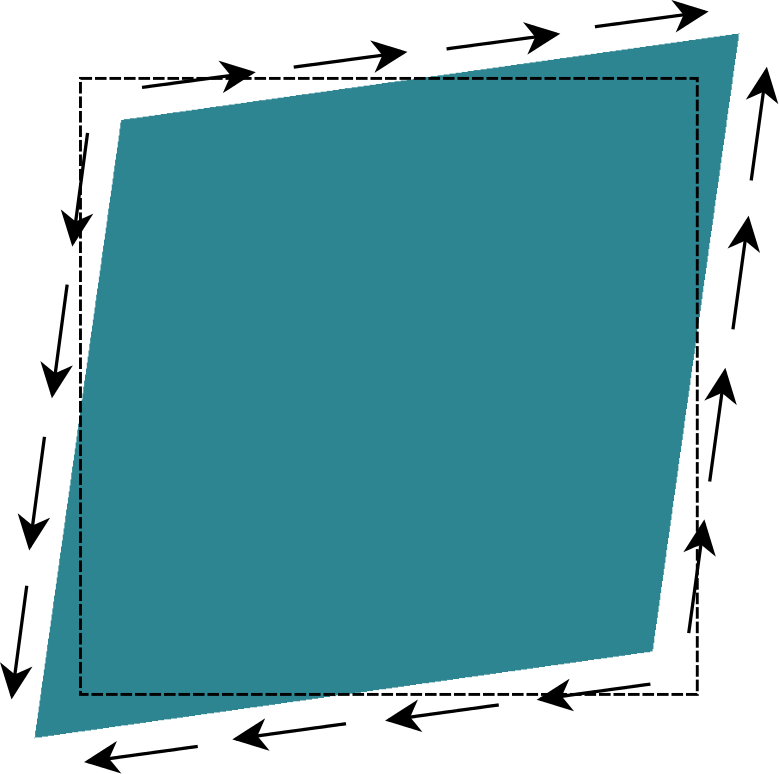}\qquad\hbox{}}%
\hfill
\includegraphics[height=0.22\textwidth]{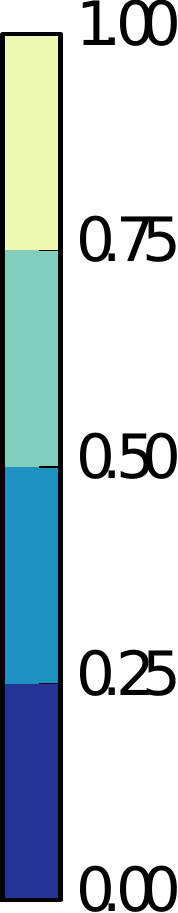}\hfill%
\subfigure[Simple shear ($\tauext$ decreases towards the surfaces)]
{\qquad\includegraphics[height=0.22\textwidth]{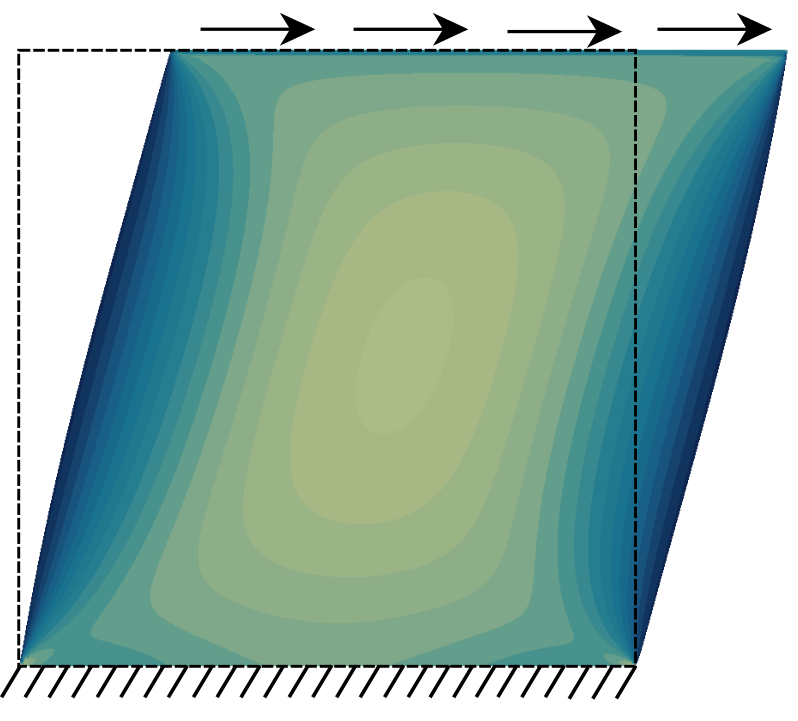}\quad\hbox{}}%
\qquad\hfill\hbox{}
\caption{\label{fig:stress_fields}%
Deformed state and scaled external elastic stress field under pure shear  and simple shear  boundary conditions. The arrows represent the distributed forces that are responsible for the respective shear deformation state.}
\end{figure}

The driving force of the time-dependent deformation of the reference model is a spatially homogeneous, external stress $\tauext$. In physical terms, this would be equivalent to a specimen of finite size that has distributed tangential forces of the same magnitude along each of its surfaces: the forces of opposite surfaces are of opposite directions such that a pure shear state (\figref{fig:stress_fields}(a)) is produced. These forces could then be applied as Neumann BCs to the FEM model.  
The magnitude of $\tauext$ is  quasi-statically increased throughout the simulation. To make the system more realistic we will then in model C obtain $\tauext$ directly from physical boundary conditions that are applied to the specimen.

\begin{table}[htp]
\centering
  \begin{tabular}{@{\quad} l || l | l @{}@{\quad}}
	                            & \bf external stress   & \bf  internal stress  \\ \hline\hline
	\bf reference model   & pure shear   & periodic Green's function    \\ \hline
	\bf  model A         & pure shear   & periodic FEM                       \\
	\bf  model B         & pure shear   & FEM with surfaces                \\
	\bf  model C         & FEM simple shear    & FEM with surfaces               \\ 
  \end{tabular}
  \caption{ \label{tab:models} Overview over the models and our approaches for computing external (\figref{fig:stress_fields}) and internal stresses.}
\end{table}

Concerning the computation of  internal stress we consider two types of boundary conditions that directly affect the elastic interactions of STs: PBCs (model A) and surfaces (model B and C).
As seen in  \figref{fig:eshelby1d}(a) free surfaces force the stresses of the interaction kernel towards zero close to the surface. This is a direct consequence of the governing equations for elasticity (compare the balance equation \eqref{eq:BE}). As a consequence, local effects of the external stress become more pronounced, while the influence of non-local interactions becomes weaker.
 
The algorithmic set-up of the FEM model is --- apart from computing the internal strains and stresses --- identical to the set-up of the reference model, which guarantees that deviations can easily be analysed.  In particular, our yield criterion consists of comparing the norm of the local shear stress with the local yield stress; we do not use any tensor-based yield criterion as e.g., based on the Mises yield stress. The yield stresses are taken for each element from a random uniform distribution between 0 and 1. Upon yielding a plastic increment (i.e., eigenstrain) of magnitude $0.2$ is added to the respective element (the shear transformation of the element).  The simulations are run until a strain of $4.0$ is reached, which is the value when the system begins to flow on average. Note that all values are dimensionless scaled quantities. We assume throughout that all systems are stress-driven and deform in plane-stress mode (i.e., the two-dimensional specimen is assumed to be thin as compared to the other directions).

\subsection{System behaviour under idealized, pure shear conditions (model A and B)}

We now consider only models A and B (Table \ref{tab:models}), subjected to the homogeneous external shear stress (\figref{fig:stress_fields}(a)), and compare the finite size scaling, the avalanche size distribution and plastic shear patterns with the reference model.

\subsubsection*{Finite size scaling}
\label{sec:FiniteSizeScaling_PBCvsFree}
\begin{figure}
	\centering
	\subfigure[Mean yield stress]{%
	\includegraphics[width=0.48\textwidth]{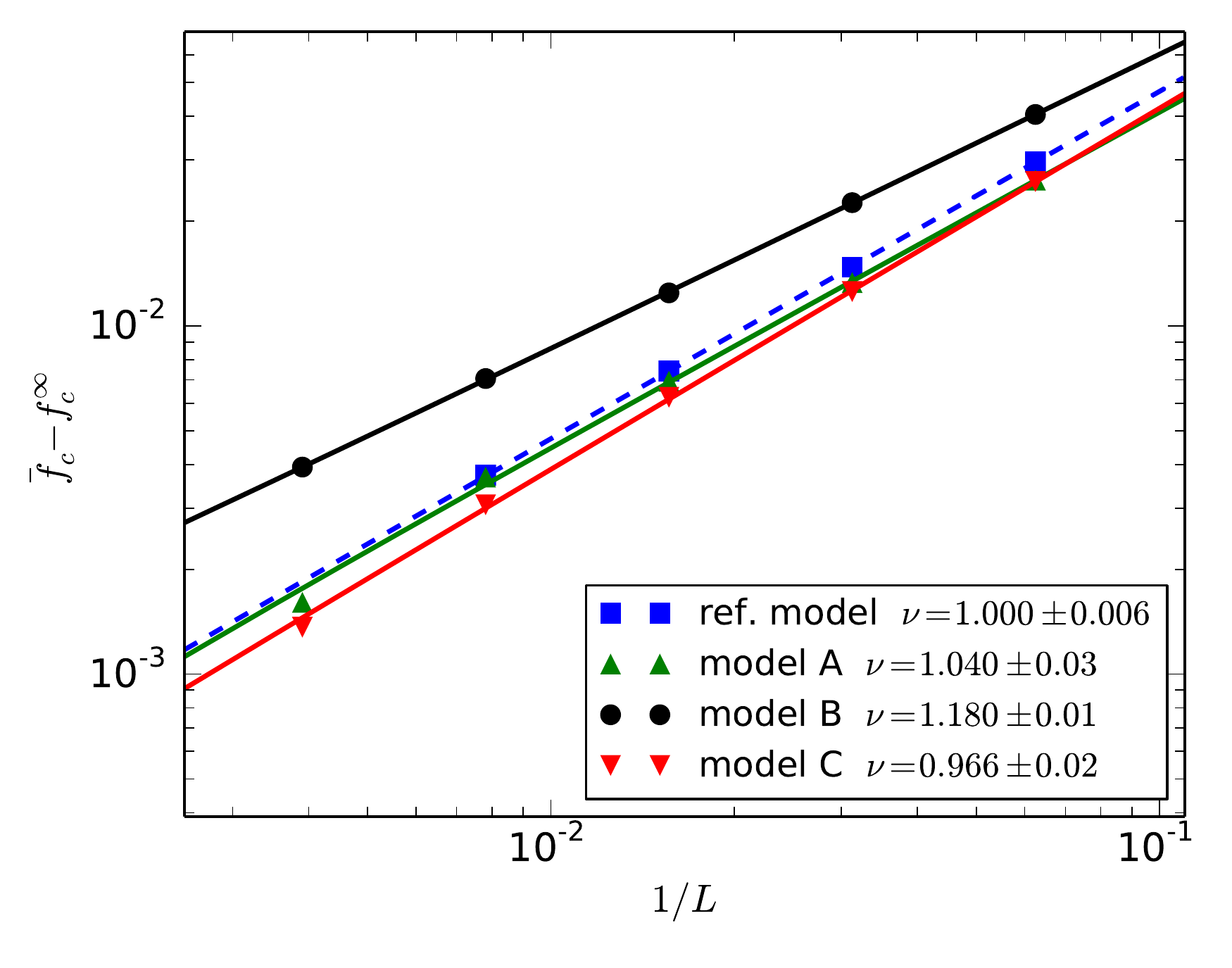}}
	\subfigure[Standard deviation of  yield stress]{%
	\includegraphics[width=0.48\textwidth]{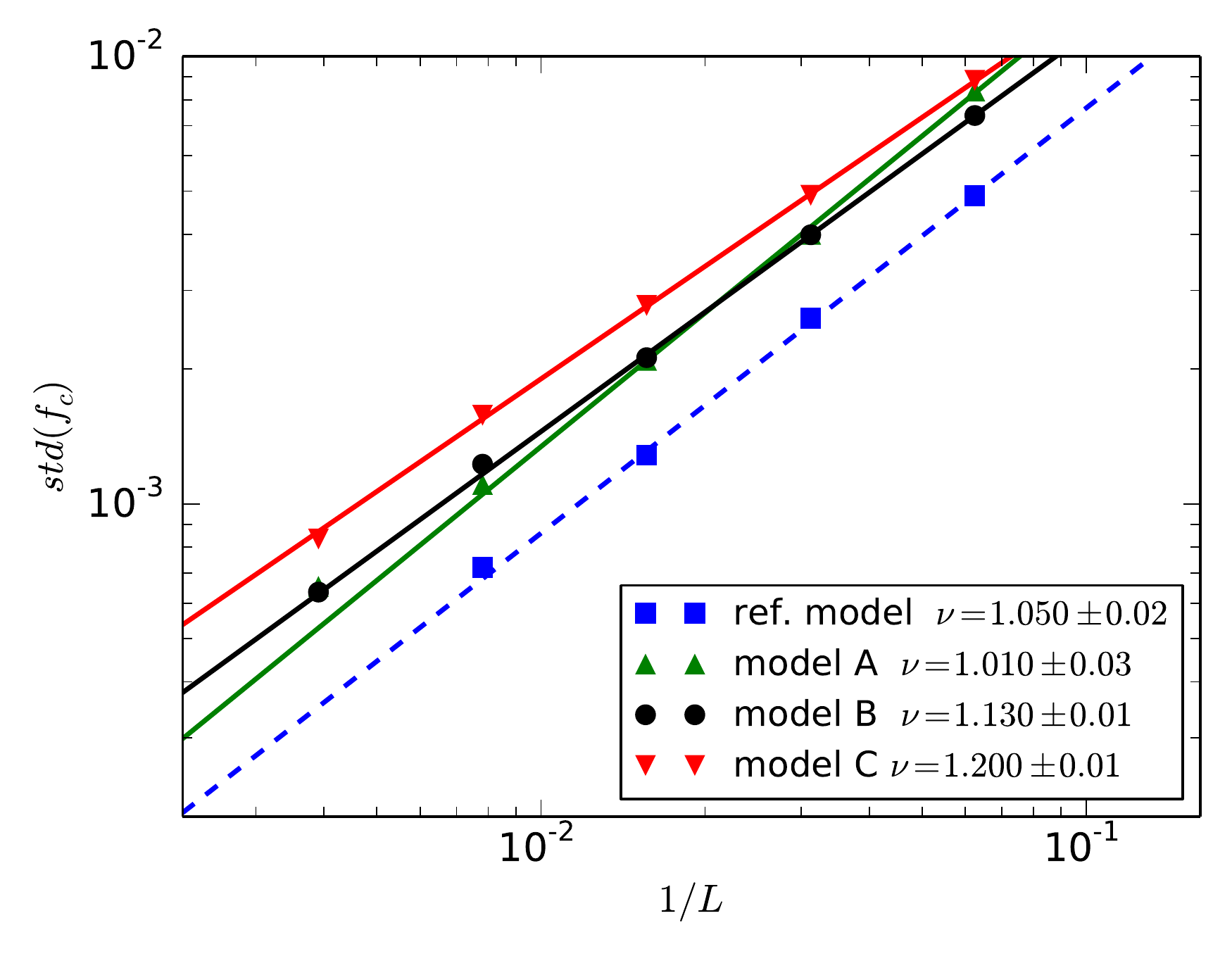}
	}
	\caption{\label{fig:yield_and_variance} 
	Finite size scaling for the mean yield stress $\bar{f_{c}}(L)$ and its standard deviation $\mathrm{std}(f_{c}(L))$. $L$ denotes the system size and the lines indicate the fit to the data.}
\end{figure}

As all phase transitions, the depinning transition is subject to finite size scaling. The measured yield stress and its standard deviation scale as:
 \begin{equation}\label{eq:fit_yield}
 \bar{f_{c}}(L) = f_{c}^{\infty}+aL^{-1/\nu} 
 \end{equation}
 
  \begin{equation}\label{eq:fit_var}
\mathrm{std}(f_{c}(L)) \propto L^{-1/\nu}
 \end{equation}
where $\bar{f_{c}}(L)$ is the mean yield stress of all simulations measured in a system of linear size $L$, $f_{c}^{\infty}$ denotes the yield stress in an infinite system and $\mathrm{std}(f_{c}(L))$ is the standard deviation of the obtained yield values.  \Figref{fig:yield_and_variance} shows the mean yield stress for system sizes of $L=16, 32, 64, 128, 256$.
We observe that  similar to the reference model (blue rectangles) the finite size scaling holds for both pure shear FEM models --- the periodic model A (green triangles) as well as for model B with free surfaces (black circles) --- and as the system becomes larger yielding occurs on average at lower stresses and simultaneously the variance decreases. Fitting the data with \eqref{eq:fit_yield} and \eqref{eq:fit_var} gives the finite size exponent $\nu$. If we average the exponents for  mean and for standard deviation, as both should be the same, we find  $\nu = 1.004 \pm 0.006$ for the reference model, $\nu = 1.03 \pm 0.02$ for the periodic model A  and $\nu = 1.16 \pm 0.07$ for model B with free surfaces.
The values found for the critical yield stress, i.e., the interpolation of the mean yield stress for a system of infinite size, are $f_{c}^{\infty}\approx 0.722$ for both FEM models and $f_{c}^{\infty}\approx 0.709$ for the reference model. The values for the reference model and model A ideally would be identical. We attribute the small difference between the latter models to the different numerical implementations of the model.

\subsubsection*{Avalanche size distributions}
\label{sec:avalanches_PBCvsFree}

\begin{figure}[htb]
 	\centering
 	\subfigure[FEM model under pure shear with PBCs (A) and free surfaces (B)]{%
 	\includegraphics[width=0.49\textwidth]{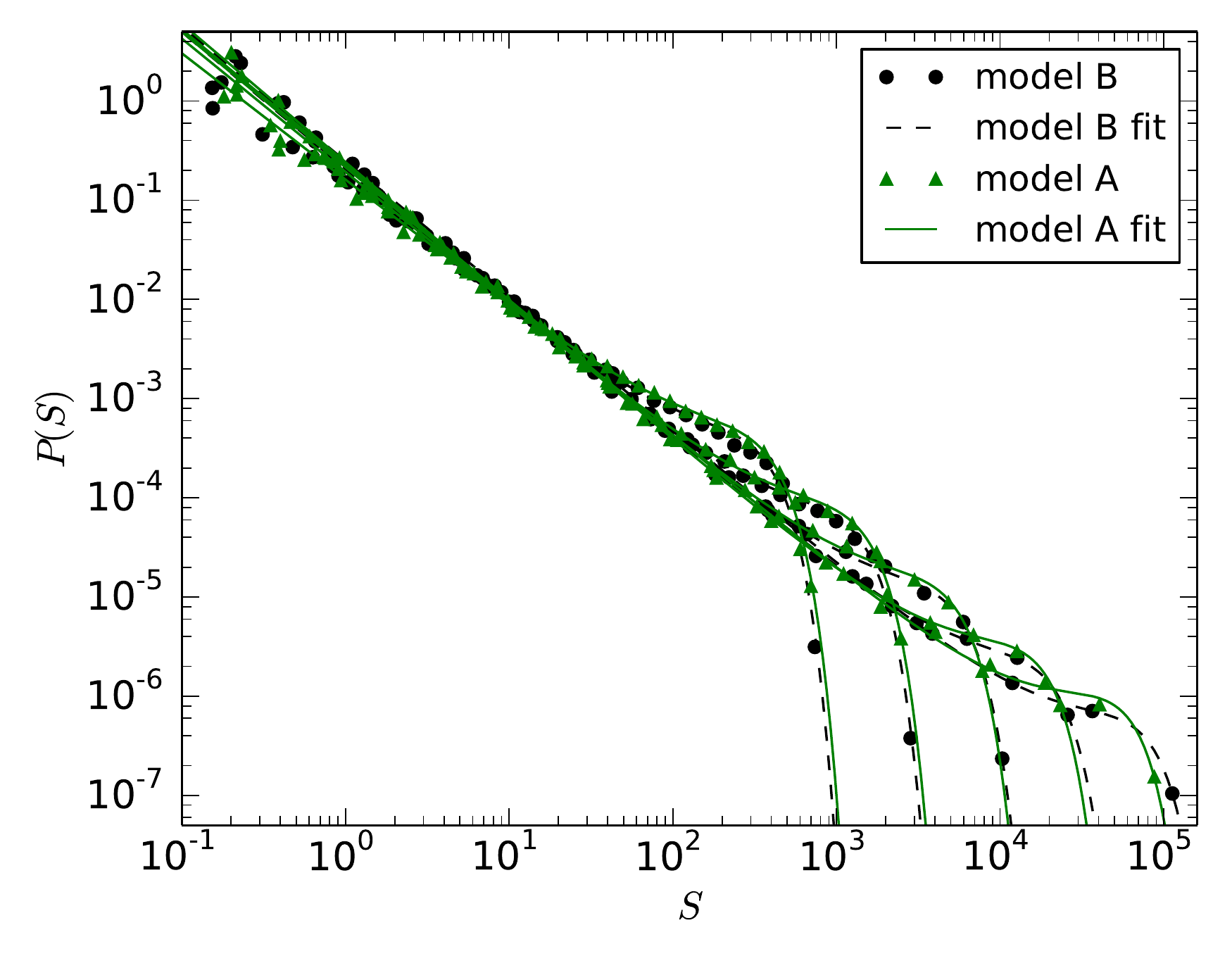}}\hfill
 	\subfigure[FEM model with surfaces under pure shear (B) and simple shear (C)]{%
 	\includegraphics[width=0.49\textwidth]{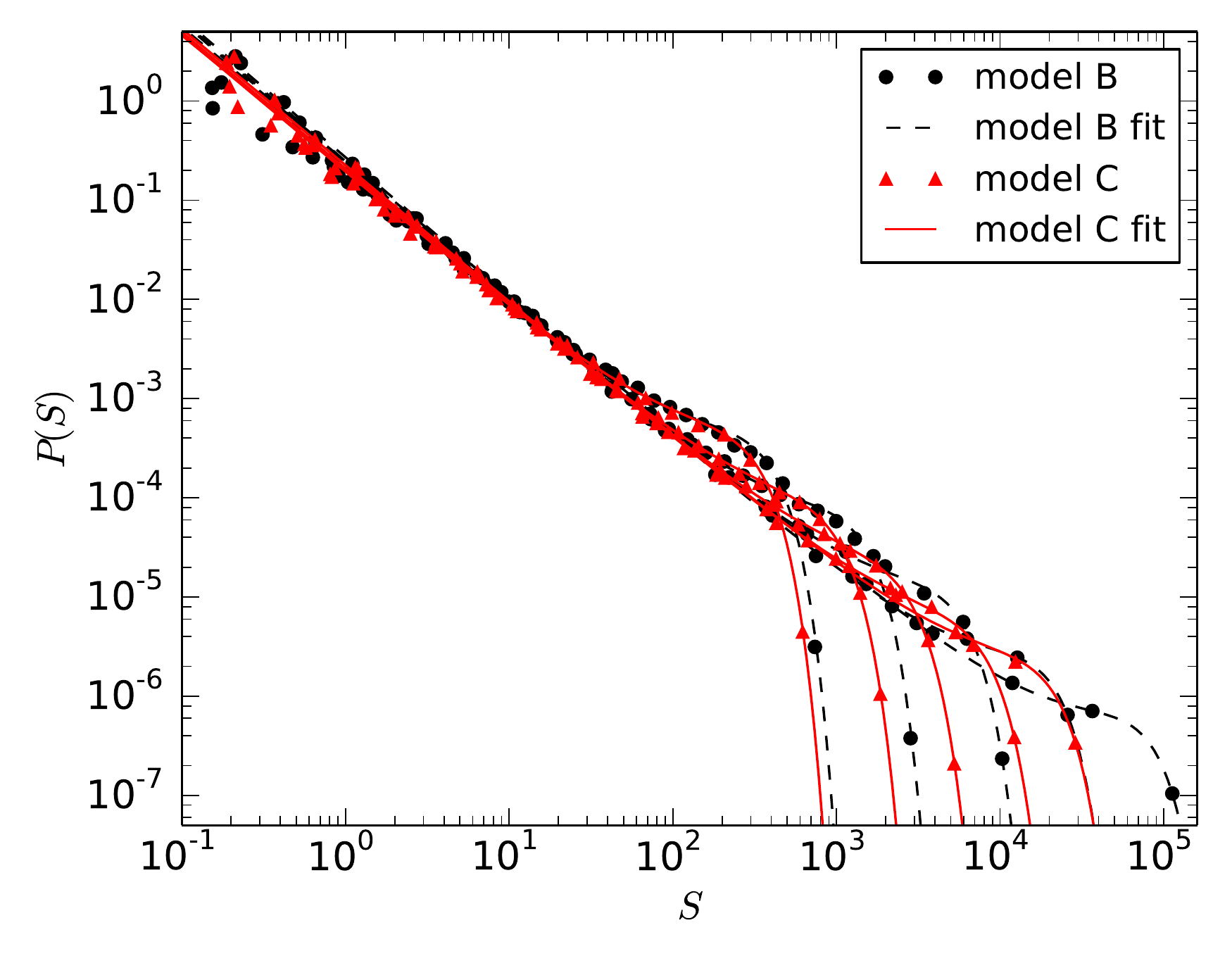}}
 
 	\caption{\label{fig:distributions} 
 	FEM avalanche size distributions and respective fits, near plastic yield as a function of system size (16, 32, 64, 128, 256 starting from the left most pair of curves) and for all three FEM models (shown as data points). The distribution is fitted to \eqref{eq:fit_aval} from which the exponents $\tau$ and $D$ are obtained. }
\end{figure}

We now compare statistics of avalanche size distributions for the two pure shear FEM models A and B with the reference model. Initially, the system is in equilibrium until the necessary stress  to trigger a first  plastic event (ST) is reached. This causes an avalanche of further plastic events together with a stress redistribution until the system reaches stress equilibrium again. The size of an avalanche is defined as the accumulation of plastic strain  in between two successive equilibrium states. 
Analysing FEM simulations near the critical point we find that the avalanches show a power-law distribution with a cut-off that features a distinct 'bump' (\figref{fig:distributions}(a)). The distributions appear to be independent of the boundary conditions (differences seen for the $L=256$ systems are likely a result of the relatively small number ($\sim200$) of realizations for that system size, rather than a real difference).

\begin{figure}
	\centering
	\subfigure[]{%
	\includegraphics[width=0.49\textwidth]%
	{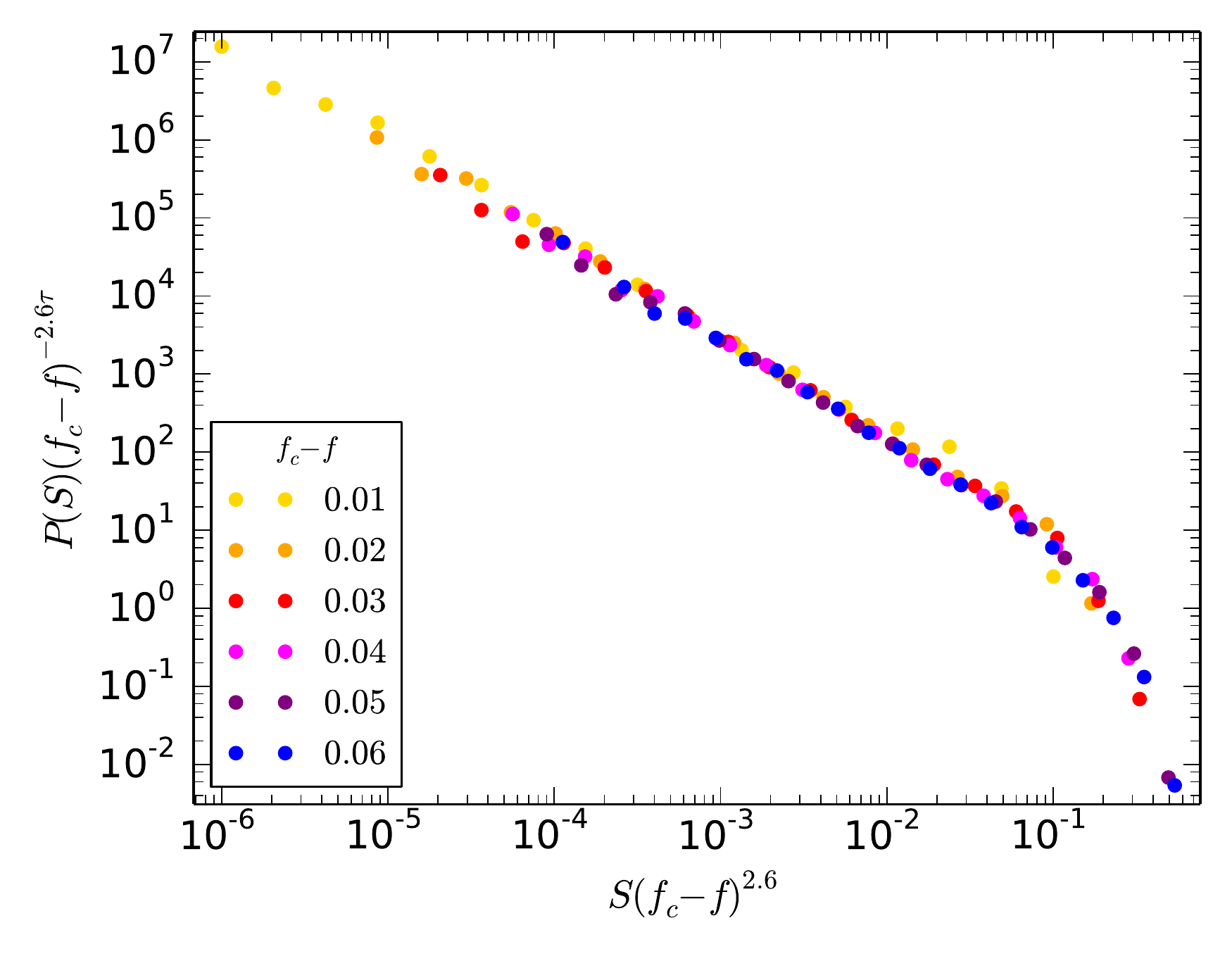}
	}
	\subfigure[]{%
		\includegraphics[width=0.49\textwidth]{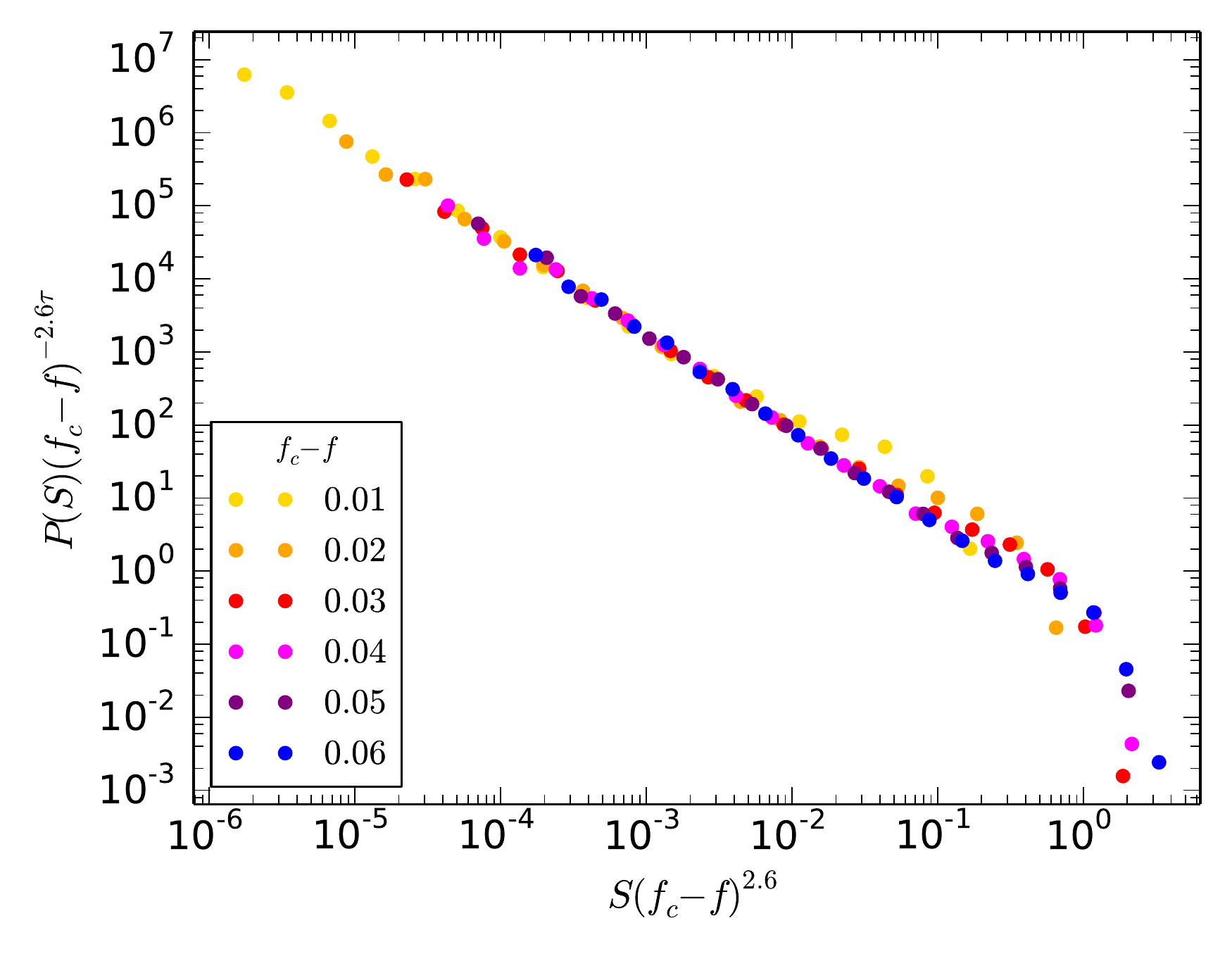}}
	\caption{\label{fig:collapse_stress} 
	Collapsed avalanche distributions according to Eq.~\ref{eq:collapse_stress} for a system of size $256\times 256$ for different stress values near plastic yield. The cutoff of the distribution, $S_0$, scales as $S_0 \propto (f_{c}-f)^{-2.6}$ under pure shear (model B) (a) and as $S_0 \propto (f_{c}-f)^{-2.6}$ under simple shear (model C) (b).}
\end{figure}

As in Ref~\cite{Budrikis2013}, we fit these distributions with the functional form:
 \begin{equation}\label{eq:fit_aval}
 P(S)=c_{1}S^{-\tau}\exp(c_{2}S-c_{3}S^{2})
 \end{equation}
For a system size of $256\times256$, the pure shear FEM models A and B both yield an exponent of $\tau = 1.36 \pm  0.01$ (\figref{fig:distributions}(a)), which agrees with the value $\tau=1.342\pm0.004$ found for the reference model in \cite{Budrikis2013}. The upper tail of the avalanche size distribution can be characterized by a cutoff $S_{0}$, which scales with distance from the critical point as 
\begin{equation}\label{eq:collapse_stress}
 S_{0} \propto (f_{c}-f)^{-1/\sigma}.
\end{equation}
We measure $1/\sigma$ by measuring the integrated exponent of the avalanche size distribution, that is, the exponent describing the power law part of the distribution when all avalanches are counted, rather than at criticality, which is given by $\tau+\sigma$. We find $1/\sigma=2.6\pm0.1$ for FEM models A and B (\figref{fig:collapse_stress}(a)), which is slightly larger than the value $1/\sigma\approx2.3$ found previously for the reference model~\cite{Budrikis2013}. However, as noted in that work, the measured value of $1/\sigma$ can depend quite strongly on the exact details of localization of plastic strain.

\begin{figure}
	\centering
	\subfigure[]{%
	\includegraphics[width=0.49\textwidth]%
	{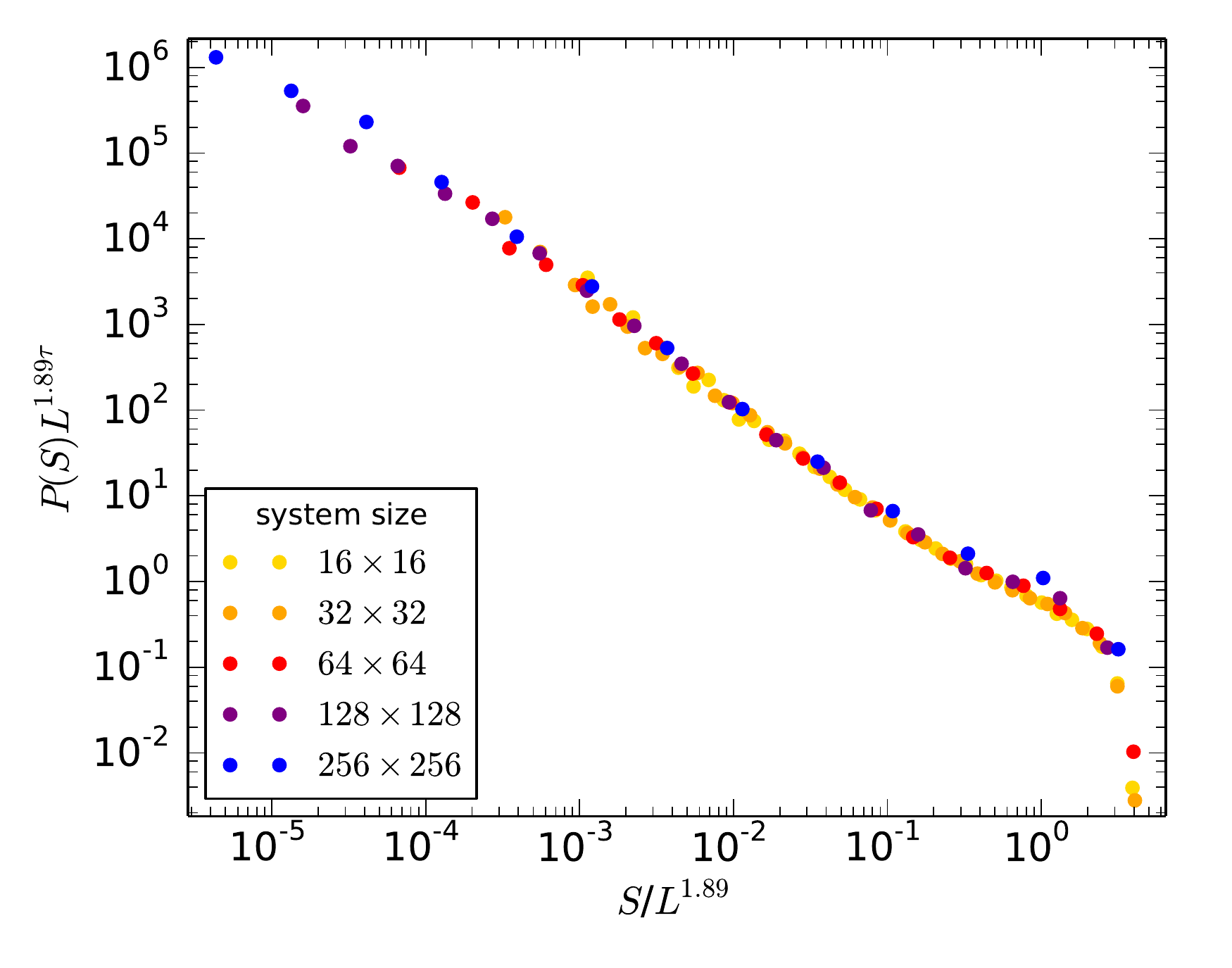}
	}
	\subfigure[]{%
		\includegraphics[width=0.49\textwidth]{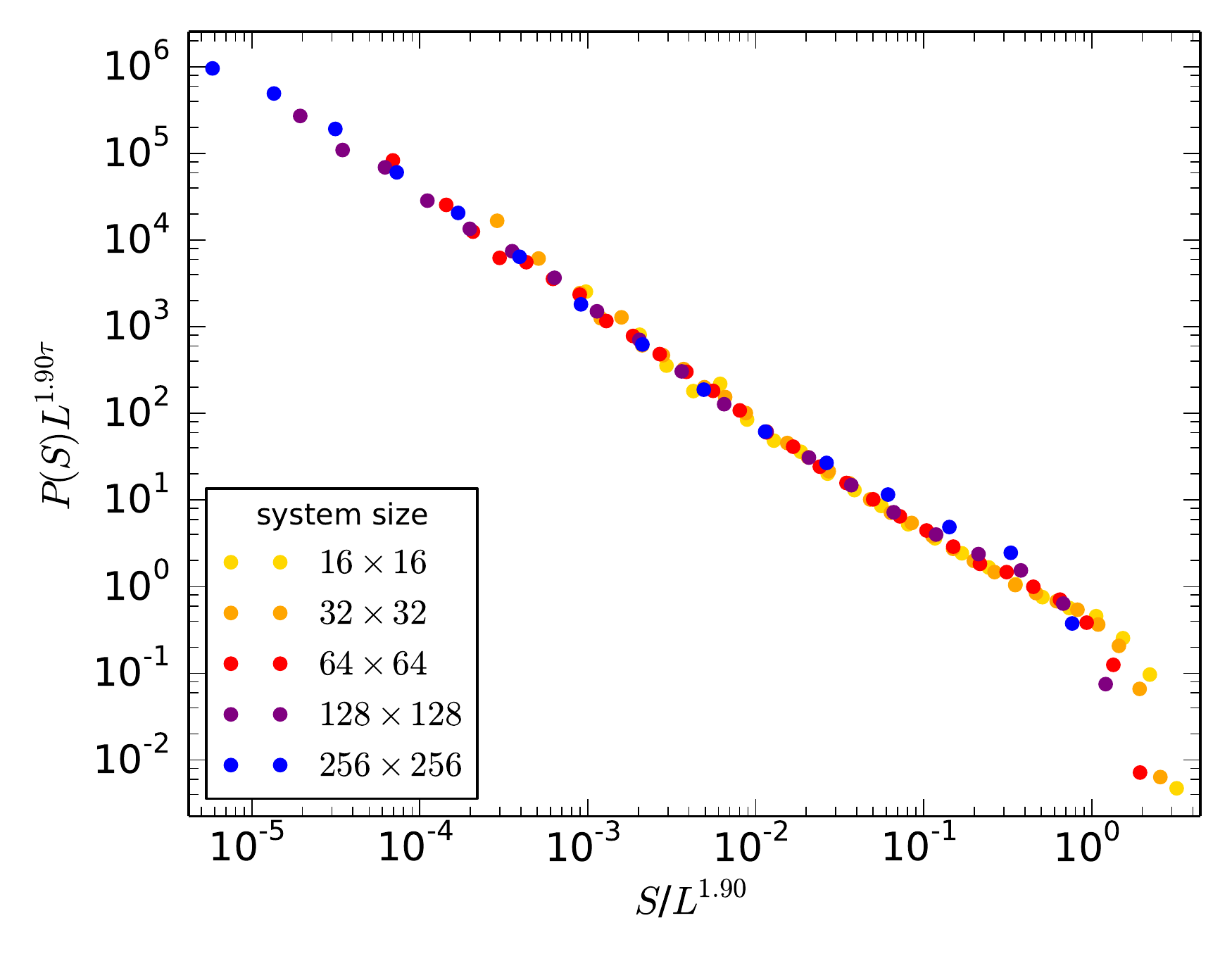}}
	\caption{\label{fig:collapse_L} 
	Collapsed avalanche distributions according to \eqref{eq:collapse_L} near plastic yield. The cutoff of the distribution, $S_0$, scales as $S_0 \propto L^{1.89}$ under pure shear (model B) (a) and as $S_0 \propto L^{1.90}$ under simple shear (model C) (b).}
\end{figure}

\begin{figure}
	\centering
	\includegraphics[width=0.49\textwidth]%
	{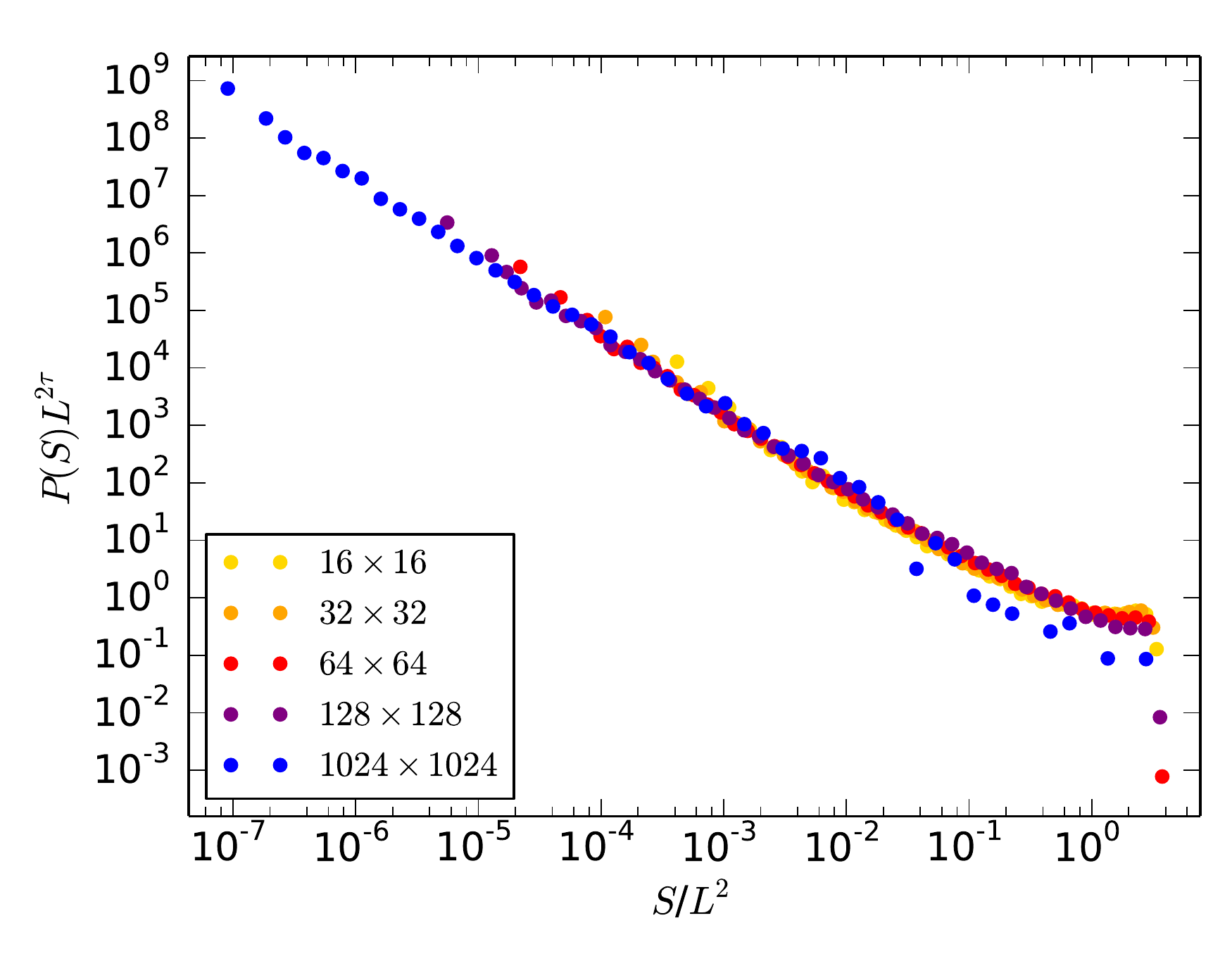}
	\caption{\label{fig:collapse_L_ref_model} 
	Reference model: near plastic yield, the cutoff $S_0$ of the avalanche size distribution scales as $S_0 \propto L^2$, as evidenced by scaling collapse of the distributions.}
\end{figure}

At the yield point, the cutoff of the avalanche size distribution depends on system size $L$ as
\begin{equation}\label{eq:collapse_L}
 S_{0} \propto L^{D}.
\end{equation} 
In our FEM simulations, we measure $D$ by taking $S_0=\sqrt{c_3}$, where $c_3$ is the fitting parameter of \eqref{eq:fit_aval}. We find $D=1.87\pm0.01$ for model A and $D=1.89\pm0.02$ for model B. Data collapse for this scaling is shown in \figref{fig:collapse_L}. These values are broadly in agreement with the value $D\approx2.0\pm 0.1$ obtained by scaling collapse for the reference model, as shown in \figref{fig:collapse_L_ref_model}.

\subsubsection*{Localization of plastic strain}
\label{patterns_PBCvsFree}

\begin{figure}[ht]
\centering
 \subfigure[Typical strain pattern for model A]%
 {\includegraphics[width=0.23\textwidth]{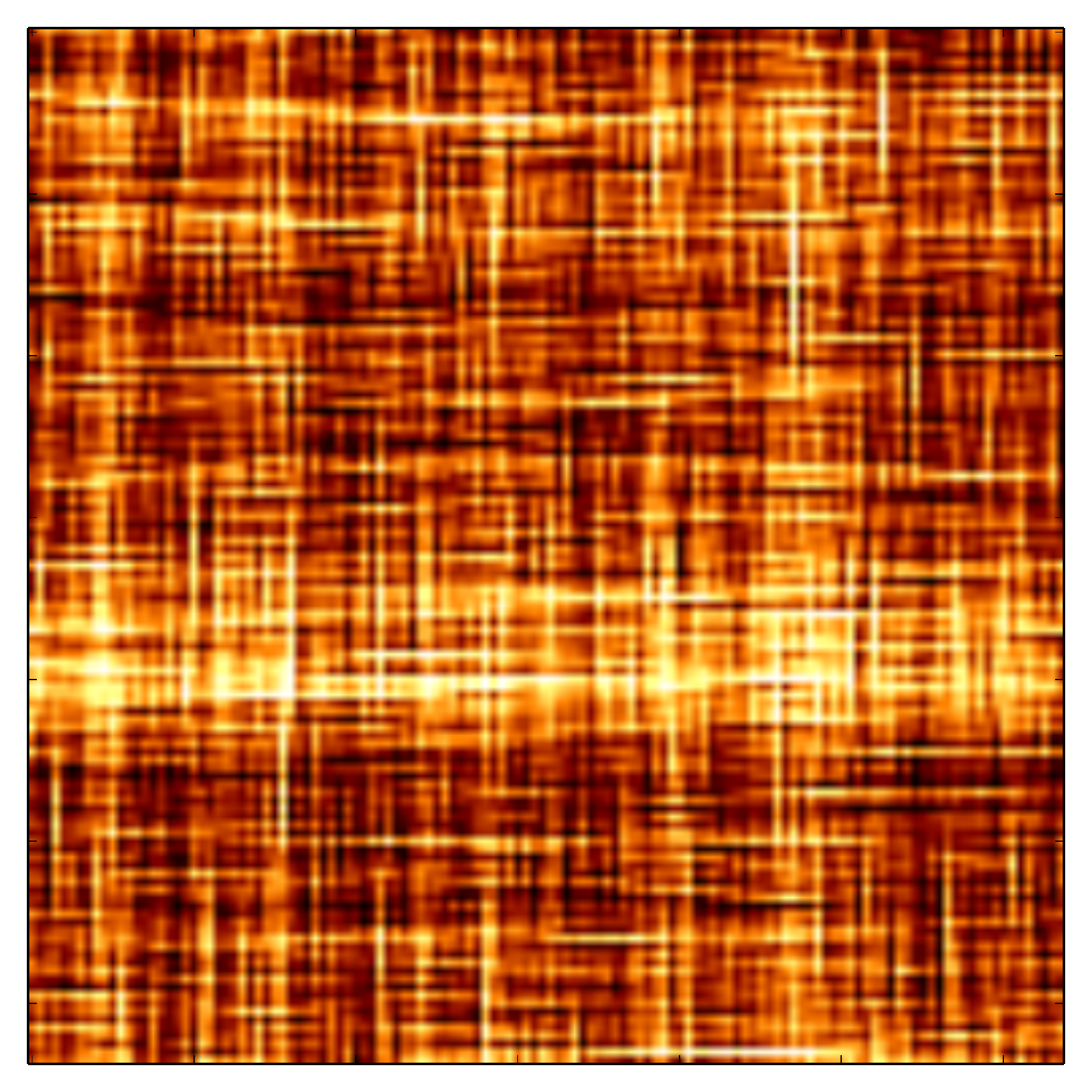}}%
\hfill%
\subfigure[Average strain for model A]
{\includegraphics[width=0.23\textwidth]{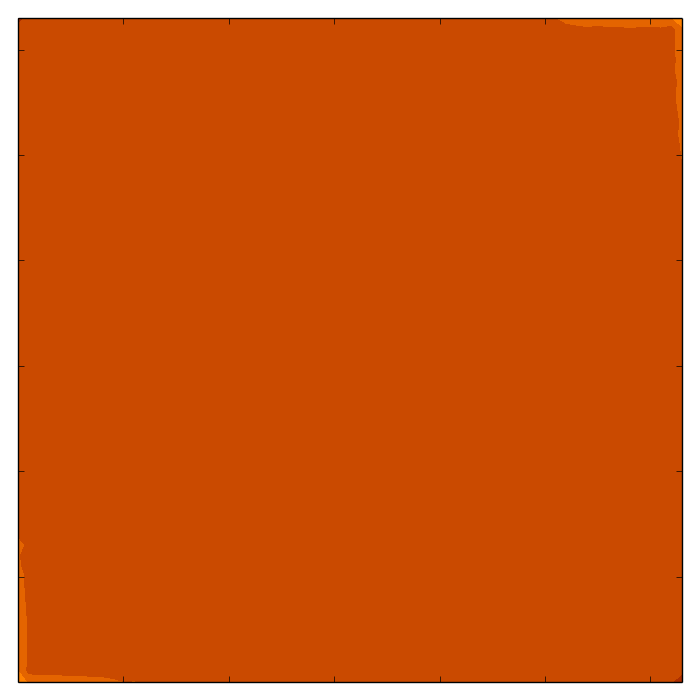}}%
\hfill%
\subfigure[Typical strain pattern model B]{\includegraphics[width=0.23\textwidth]{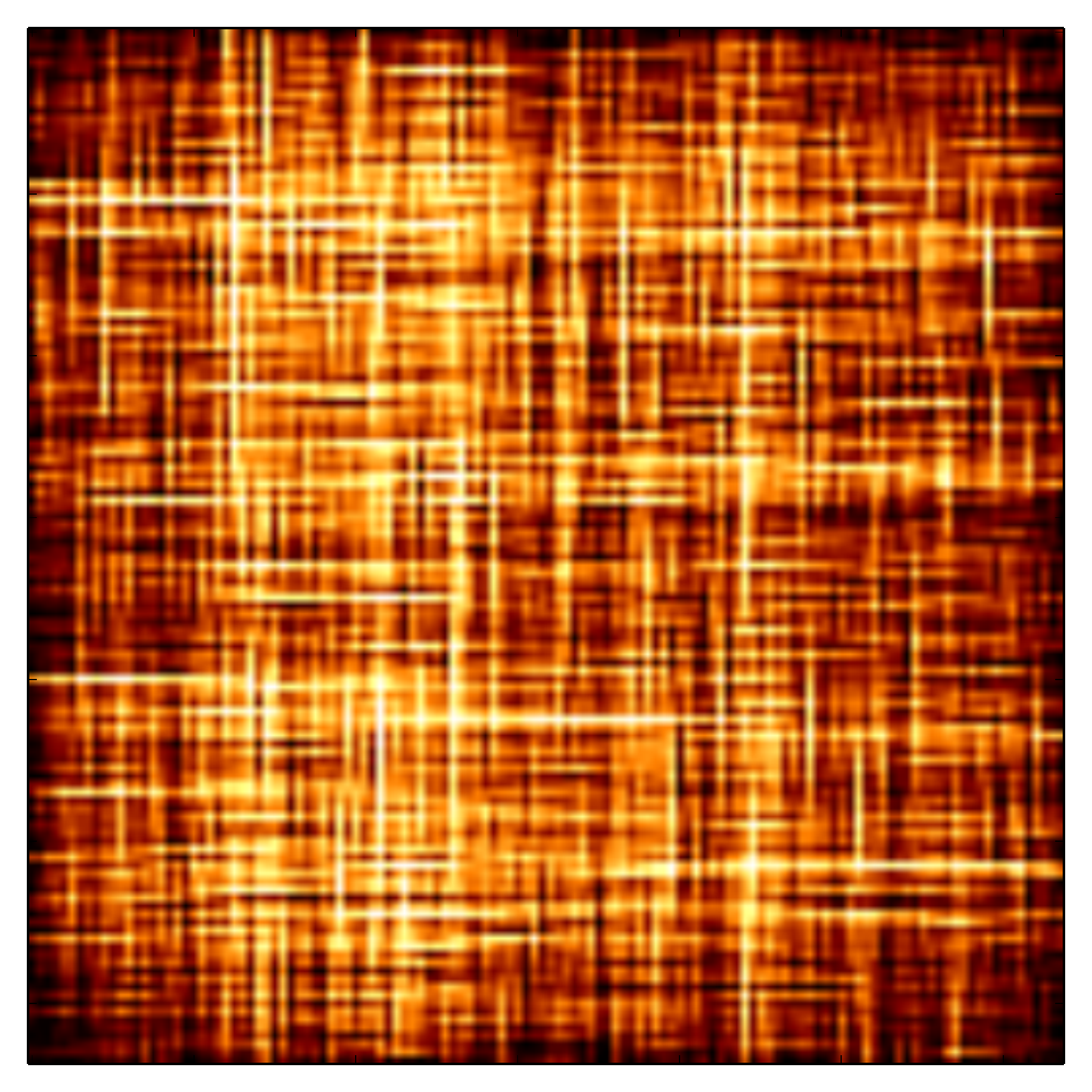}}%
\hfill%
\subfigure[Average strain for model B]%
{\includegraphics[width=0.23\textwidth]{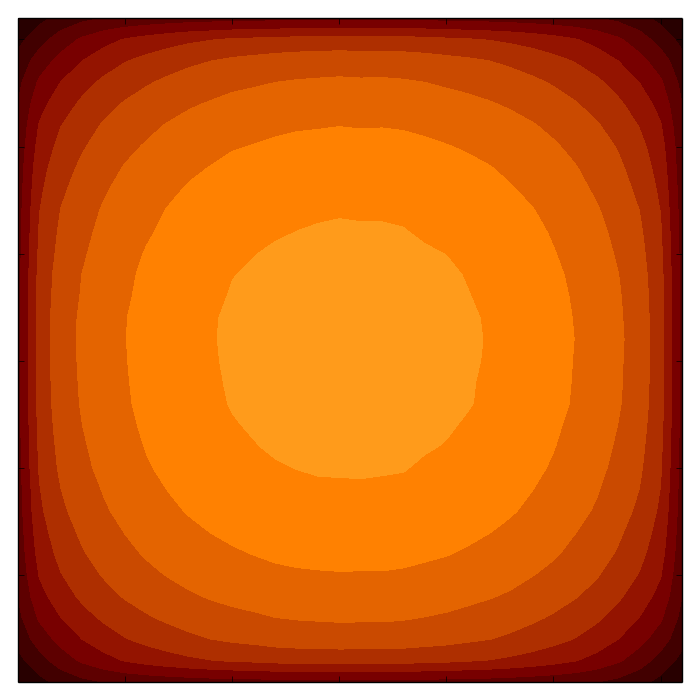}}\hfill%
\,
\includegraphics[height=0.23\textwidth]{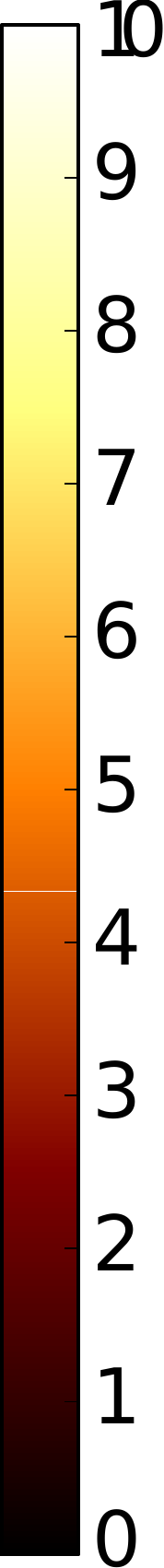}%
\caption{\label{fig:strain_maps}%
Plastic strain patterns for the pure shear models A and B. The average strain maps (b) and (d) were obtained as ensemble averages over 6000 realizations.}
\end{figure}

As a consequence of the elastic interaction between the STs, we can observe how shear bands appear, as shown in \Figref{fig:strain_maps}. 
To analyse the effect of free surfaces on plastic strain localization we ensemble average the plastic strain map of approximately 60000 simulations. We observe that for the periodic pure shear model A (\figref{fig:strain_maps}(b))  the average strain map is constant. The average plastic strain value of $4.0$ corresponds to the plastic strain at which we terminate our simulations. If we compare to the non-periodic pure shear model B, we observe that the strain distribution is no longer constant but rather decays towards the surfaces (\figref{fig:strain_maps}(d)). In particular, the value near  the surfaces is lower than the average of $4.0$, and towards the center the value is above the average. Since the surfaces can deform more freely the surface stresses are reduced. Hence, the plastic strain is lower which is compensated by an increase in the inner region. On the other hand, the free surface boundary conditions do not have a visible effect on the strain patterns within the bulk. For example, the typical width of strain localizations remains the same in \figref{fig:strain_maps}(a) and (c), indicating that the localization width is governed by  short-range interactions, which is in agreement with Ref.~\cite{Budrikis2013}.

\subsection{System behaviour under heterogeneous, simple shear conditions (model C)}
\label{sec:PurevsSimple} 
So far we have only utilized the FEM for computing the resulting stress that are caused by an eigenstrain distribution. As introduced above, FEM can easily  handle loading situations that are more realistic than the uniform pure shear loading that we used in the previous section. Therefore, we now use FEM to compute both the internal and external stresses (Table \ref{tab:models}). The external stresses arise from simple shear boundary conditions (\figref{fig:stress_fields}(b)), in which the bottom edge of the system is fixed stationary and the top is fixed vertically but can move horizontally under the effect of an applied lateral traction force. The left and right surfaces remain free of tractions or constrictions (free surfaces as in models A and B). The resulting external stress field is shown in \figref{fig:external_field_simple_shear} when the applied force is $0.6$, a value at which a typical system yields;  all other parameters of the simulations remain the same as those used for the pure shear simulations in \secref{sec:PBCvsFree}. Obviously, the external stress resulting from the simple shear situation exhibits significant deviations from the constant, pure shear stress field. Most notably, the stress field has strong gradients, as a result of the continuum mechanical balance equation at a free surface. We emphasize that the stress field regardless the size of the system never exhibits a plateau of constant stress. How this impacts the scaling and shear banding behaviour as compared to our previously studied models will be analysed subsequently.

\subsubsection*{Finite size scaling}
\label{FiniteSizeScaling_PureVsSimple}
Analysing the yield stress distributions under simple shear loading, we obtain again the mean yield stress and standard deviation. Both quantities follow a power law (\figref{fig:yield_and_variance}) similar to that found for the pure shear models. Averaging the exponents for the mean and for the standard deviation, we find values of $\nu = 1.15 \pm 0.09$ for the simple shear model C and $\nu = 1.16 \pm 0.07$ for the pure shear model B.
The values found for the critical yield stress are $f_{c}^{\infty}\approx 0.662$ for model C, compared to $f_{c}^{\infty}\approx 0.722$ for the system under pure shear as seen in \secref{sec:FiniteSizeScaling_PBCvsFree}. The difference is large and  suggests that the macroscopic yield stress is strongly dependent on the loading condition. This is in accordance to what is known from experimentally tested samples and which motivated the introduction of different measures for the `equivalent stress' as e.g., the \emph{von Mises} stress.

\subsubsection*{Avalanche size distributions}
\label{sec:avalanches_PureVsSimple}
Analysing the avalanche distributions in the same way as in \secref{sec:avalanches_PBCvsFree}, we observe that both systems exhibit an avalanche distributions with approximately the same slope in the power law regime, as shown in \figref{fig:distributions}(b), with measured exponent $\tau = 1.32 \pm  0.02$. However, the stress dependence of the cutoff of the avalanche distributions has a clear dependence on loading conditions, and for simple shear we measure $1/\sigma\approx 2.6$. Furthermore, the cutoff is also found to scale with system size as $L^{D}$ with $D=1.90\pm0.01$ (shown in \figref{fig:collapse_L}). These exponents should be compared to the pure shear values $\tau = 1.36 \pm  0.01$, $1/\sigma= 2.6\pm0.1$ and $D=1.89\pm0.02$  (cf. \secref{sec:avalanches_PBCvsFree}). In other words, the exponents are unaffected by the loading conditions.

\subsubsection*{Localization of plastic strain}
\label{patterns_PureVsSimple}

\begin{figure}[ht]

	\hbox{}\hfill
	\subfigure[Simple shear external stress field ]{%
	$\vcenter{\hbox{\includegraphics[width=0.24\textwidth]{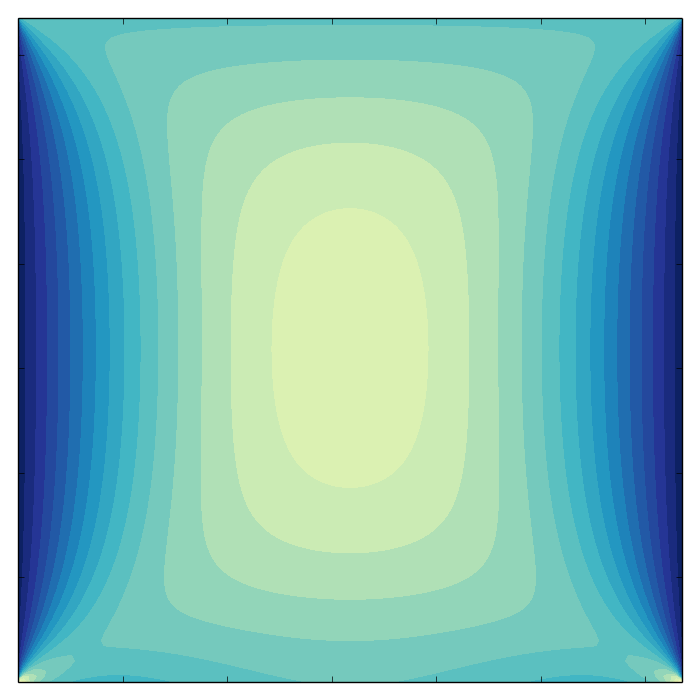}}}$\;%
	$\vcenter{\hbox{\includegraphics[height=0.23\textwidth]{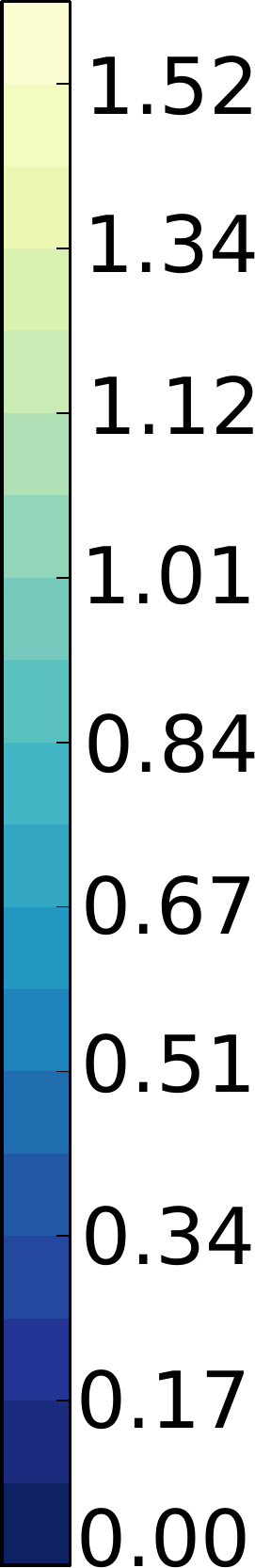}}}$%
	\label{fig:external_field_simple_shear} 	}
	\hfill	
	\subfigure[Typical plastic strain pattern (model C)] {$\vcenter{\hbox{\includegraphics[width=0.24\textwidth]{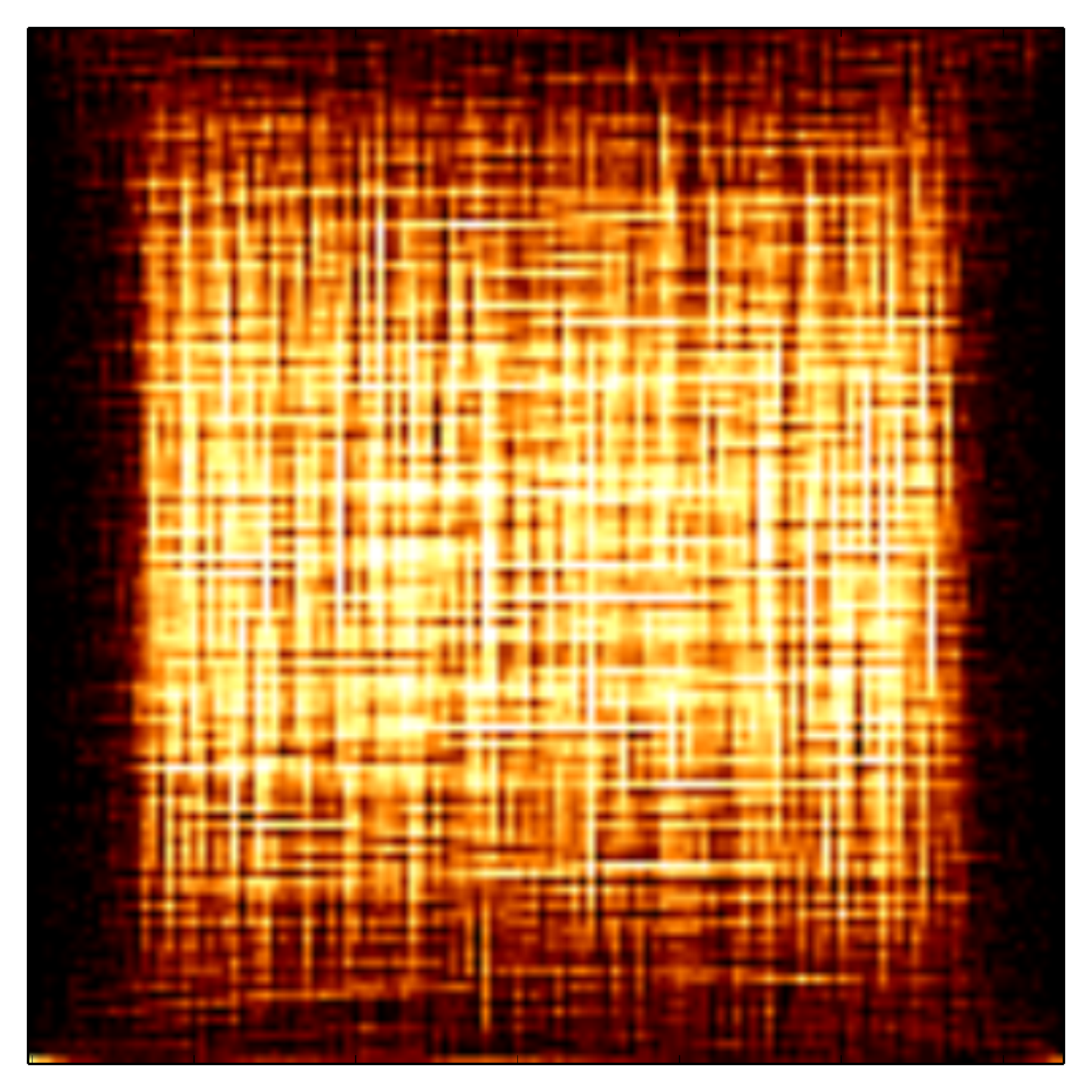}}}$}\hfill%
	\subfigure[Averaged plastic strain pattern (model C)]{%
		$\vcenter{\hbox{\includegraphics[width=0.24\textwidth]{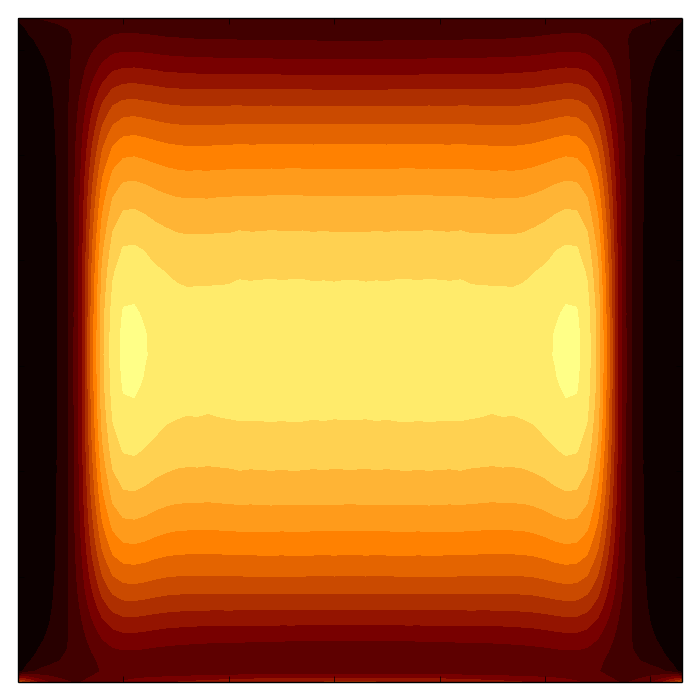}}}$	}\;
    	$\vcenter{\hbox{\includegraphics[height=0.23\textwidth]{figures/color_scale_strain.pdf}}}$
	\hfill\hbox{}
   
    \caption{\label{fig:strain_maps_simple_shear} Strain patterns for the simple shear simulation (model C) where the system is loaded by a lateral force on the top face. The average plastic strain pattern are obtained from $\approx$ 20000 realizations.}
\end{figure}

The plastic strain localization under simple shear loading, shown in \figref{fig:strain_maps_simple_shear}, exhibits a non trivial localization pattern. In this case, the system is affected simultaneously by the effects of the free surfaces at the left and right faces and by prescribed vertical displacements at the top and bottom, which all affect the internal stresses. Additionally, the strongly heterogeneous external stress field resulting from the lateral forces together with the aforementioned boundary conditions promotes plastic activity in only some regions of the system. From the symmetry of the external shear stress field  (\figref{fig:external_field_simple_shear}), a higher strain localization would be expected at the center of the system. However, this is in contrast with the obtained strain patterns. Additionally,  from the average plastic strain pattern (\figref{fig:external_field_simple_shear}(c)), two maxima can be observed near the left and right vertical faces. The rest of the system behaves as expected: small plastic activity at the top and bottom regions (where the external stress is small and the interaction of the STs tends to zero), and almost zero plastic activity at the left and right surfaces (where both the external stress and STs interaction tends to zero).

\section{Discussion and Conclusions}
\label{sec:results}
The effect of boundary conditions on strain avalanches and localization is an important problem that must be addressed to make rigorous links between statistical models and experiments on amorphous materials. While some progress has been made with analytical calculations~\cite{Picard2004,Nicolas2014}, it is clear that numerical methods are also needed to fill the gap. Along these lines, we have presented simulations that take advantage of finite element tools to tackle these problems.
Our finite element simulations reproduce the behaviour of the reference system when we assume appropriate boundary conditions, which is confirmed by comparing model A (PBCs for the internal and a pure shear stress state from the external stresses) with the reference model. By changing boundary and loading conditions, we extend the model beyond what is possible in a simple lattice model. 
\begin{table*}[htp]
\centering
  \begin{tabular}{@{\quad} l | l | l | l | l @{}@{\quad}}
    \hline
                 & $\nu$ &   $\tau$ & $D$ & $1/\sigma$  \\ \hline\hline
        ref. model & $1.004\pm0.006$  & $1.342\pm0.004$, & $\sim 2\pm 0.1$   & $2.3\pm0.05$, \\
                     &                  & cf. Ref.\cite{Budrikis2013} & &  cf. Ref. \cite{Budrikis2013}\\       \hline
        model A & $1.03\pm0.02$    & $1.36\pm0.01$    & $1.87\pm0.01$   & $2.6\pm0.1$ \\
        model B & $1.16\pm0.07$    & $1.36\pm0.01$    & $1.89\pm0.02$   & $2.6\pm0.1$ \\
        model C & $1.15 \pm 0.09$  & $1.32\pm0.02$    & $1.90\pm0.02$ & $2.6\pm0.1$ \\         \hline
  \end{tabular}
  \caption{ \label{exponents_list} Critical exponents for the reference model, model A (PBCs and pure shear deformation), model B (free surfaces and pure shear deformation) and model C (surfaces and simple shear deformation).}
\end{table*}

Boundary conditions affect the long range part of stress fields generated by inclusions, in particular, how they deviate from power-laws (as illustrated in \figref{fig:eshelby1d}). Our models A and B differ only in boundary conditions --- periodic and free surfaces, respectively --- and can therefore be compared to examine the effects of changing boundary conditions.
%
Free surfaces require the stresses to drop to zero on the system boundary, which leads to a decrease in plastic deformation near the edges (\figref{fig:strain_maps}(d)), while the deformation in the bulk shows the same characteristic patterns as for PBCs (\figref{fig:strain_maps}(a) and \figref{fig:strain_maps}(c)). This behaviour is even more pronounced for the simple shear system (model C) due to the superposition with the non-homogeneous external stress field.

We have also examined the effect of boundary conditions on critical exponents. In the transition from periodic systems (reference model and model A) to systems with surfaces (model B and C), we find that changing only boundary conditions has little effect on critical exponents, as seen in Table~\ref{exponents_list}. We can observe a small increase in the value of the exponent $\nu$ associated with the different interaction kernels for periodic and non-periodic systems (\figref{fig:eshelby1d}(a)).

%
%
Changing loading conditions while keeping surface boundary conditions does not affect the interaction kernel, so a comparison of model B (pure shear loading) and model C (simple shear loading) can be used to test whether external shear stress distribution has an effect on critical exponents. As with changes in boundary conditions, we find little effect. We conclude therefore that universal behaviour measured in periodic lattice models such as our reference model can be expected also in more realistic loading conditions, except for a small change in the exponent $\nu$ related to the existence of surfaces. On the other hand, the localization of plastic strain is determined not only by the range and anisotropy of interactions~\cite{Martens2012}, but also boundary and loading conditions. Therefore, care should be taken before drawing strong conclusions from strain localization observed in simulations with a set up similar to the reference model.



\section*{Acknowledgments}
The authors would like to thank Michael Zaiser for fruitful discussions. S.S. and D.C. gratefully acknowledge financial support from the German Research Foundation (DFG) under the DFG Research Group FOR 1650 `Dislocation-based Plasticity', contract Sa-2291/1-1; Z.B. and S.Z. are supported by the European Research Council through the Advanced Grant 2011 SIZEFFECTS.  S.Z. acknowledges support
from the Academy of Finland FiDiPro progam, project 13282993.

 \section*{References}
\bibliographystyle{unsrt}
\bibliography{paper}

\end{document}

%% file: symbols_and_definitions.tex
\newcommand{\Bu}{{\boldsymbol{\mathnormal u}}}

\newcommand{\Br}{{\boldsymbol{\mathnormal r}}}
\newcommand{\superscr}[1]{\ensuremath{{}^{\rm #1}}}

\newcommand{\Bsigma }{\ensuremath{\boldsymbol\sigma}}

\newcommand{\Bve    }{\ensuremath{\boldsymbol\varepsilon}}
\newcommand{\ve     }{\ensuremath{           \varepsilon}}

\newcommand{\epsint }{\ensuremath{           \varepsilon\superscr{int}}}

\newcommand{\tauext }{\ensuremath{           \tau \superscr{ext}}}

\newcommand{\tauint }{\ensuremath{           \tau \superscr{int}}}

\DeclareRobustCommand{\calV}{\ensuremath{\mathcal V} }

\newcommand{\lint }[2]{\displaystyle\int\limits^{#1}_{#2}}

\newcommand{\totdiff}{\textrm{d}}

\DeclareMathAlphabet{\Ibb}{U}{msb}{m}{n}
\newcommand{\IC}{{\ensuremath{\Ibb C}}}

\newcommand{\IR}{{\ensuremath{\Ibb R}}}

\renewcommand{\div}{\textrm{div}}

